\documentclass[hyper]{JHEP3} 

\usepackage{axodraw}
\usepackage{epsfig}
\usepackage{amsmath}
\usepackage{cite}

\newcommand{\ba}{\begin{array}}
\newcommand{\ea}{\end{array}}
\newcommand{\bd}{\begin{displaymath}}
\newcommand{\ed}{\end{displaymath}}
\newcommand{\be}{\begin{equation}}
\newcommand{\ee}{\end{equation}}
\newcommand{\bea}{\begin{eqnarray}}
\newcommand{\eea}{\end{eqnarray}}

\def\q2 {q^2}

\def\bt{\begin{table}}
\def\et{\end{table}}

\title{Non-universal scalar masses: a signal-based analysis
for the Large Hadron Collider}

\author{Subhaditya Bhattacharya\\
         Regional Centre for Accelerator-based Particle Physics \\
     Harish-Chandra Research Institute\\
Chhatnag Road, Jhusi, Allahabad - 211 019, India\\
        E-mail: \email{subha@mri.ernet.in} }
\author{AseshKrishna Datta\\
         Regional Centre for Accelerator-based Particle Physics \\
     Harish-Chandra Research Institute\\
Chhatnag Road, Jhusi, Allahabad - 211 019, India\\
        E-mail: \email{asesh@mri.ernet.in} }
\author{Biswarup Mukhopadhyaya\\
         Regional Centre for Accelerator-based Particle Physics \\
     Harish-Chandra Research Institute\\
Chhatnag Road, Jhusi, Allahabad - 211 019, India\\ 
        E-mail: \email{biswarup@mri.ernet.in} }

\abstract{ We study the possible signatures of non-universal scalar masses 
in supersymmetry at the Large Hadron Collider (LHC). This is done,
following our recent study on gaugino non-universality, via a 
multichannel analysis, based largely on the ratios of event rates
for different final states, aimed at minimizing irregularity in the
pattern due to extraneous effects and errors. We have studied 
(a) squark-slepton non-universality, (b) non-universality in sfermion
masses of the third family, (c) the effects of $SO(10)~ D$-terms in 
supersymmetric Grand Unified Theories. After presenting an elaborate
numerical analysis of like- and opposite-sign dileptons, inclusive and
hadronically quiet trileptons as well as inclusive jet final states,
we point out specific features of the spectrum in each case, which
can be differentiated in the above channels from the spectrum for
a minimal supergravity scenario with a universal scalar mass at high scale.
The event selection criteria, and the situations where the signals 
are sizable enough for a comparatve study, are also delineated. It is found
that, with some exceptions, the trilepton channels are likely to be
especially useful for this purpose.}

\keywords{Supersymmetric Standard Model, GUT, Supersymmetry phenomenology, 
Hadronic Colliders }

\preprint{HRI-P-08-04-002\\
HRI-RECAPP-08-05}

\begin{document}

\section{Introduction} 

With the Large Hadron Collider (LHC) likely to become operative 
in the immediate future, it is of great importance to sharpen
the prediction and analysis of different types of physics beyond 
the standard model (SM). This consideration applies especially to
supersymmetry (SUSY), because (a) SUSY is one of the most frequently
explored options for new physics, and (b) a large variety of SUSY
scenarios offer themselves as candidate theories, 
often substantially different from each other in their
phenomenological implications\cite{Book,Sally,Gl,Martin}.

The much-advertised merits of SUSY, and at the same time the
concerns voiced in connection with it, lead to the expectation 
of the following features:

\begin{itemize}
\item Stabilization of the electroweak scale.

\item The existence of a cold dark matter candidate.

\item The possibility of paving the path towards 
a Grand Unified Theory (GUT).

\item Relating SUSY to some overseeing physics at the 
Planck scale, or a similarly high scale of energy.

\item Ensuring the suppression of flavour-changing
neutral currents (FCNC).
\end{itemize}

The first of these features necessitates a spectrum of new particles,
at least the gauginos, higgsinos and the third 
family of squarks and sleptons, within the
TeV scale. The second requirement is a motivator towards conserved
$R$-parity, defined as $R = (-1)^{(3B + L + 2S)}$ \cite{CDM}. 
While the ambition for GUT inspires one to envision the 
SU(3), SU(2) and U(1) gaugino
masses (together with the corresponding gauge couplings) 
as related at a high scale, the scalar (or sfermion) masses also
may be traced to some high-scale origin if physics at that scale
subsumes the low-energy SUSY scenario \cite{Martin,GUT1,GUT2}. 
And finally, the scalar spectrum
of SUSY may be subject to specific constraints if FCNC processes
are to be suppressed \cite{FCNC}. 
It is with the above requirements in view that
some simplified models of SUSY breaking are pursued, of which the
most popular one is one conserving $R$-parity and based on minimal
supergravity (mSUGRA) with universal scalar and gaugino masses at the
energy scale where SUSY is broken in a postulated `hidden sector'
\cite{Martin,mSUGRA}.
Since one is not sure whether TeV-scale physics is indeed 
dictated by an idealization of the above form, an important question
to ask is: can the departure from a scenario with universal
scalar and gaugino masses (such as in mSUGRA) be reflected in
signals observed at the LHC? 

Although this question has been explored in earlier works, the 
need of systematic analyses, based primarily on observable
signals, still remains. In an earlier study, we have investigated
the effects of departure from gaugino universality (even within
the ambit of a SUSY-GUT scenario) on
various signals at the LHC, and identified situations
where a multichannel analysis can reveal traces of such
departure \cite{Subho}. In the present work, we take up a similar
investigation of departure of the squark and slepton spectrum
 from that predicted by mSUGRA. A number of theoretical 
scenarios have already been investigated in this connection.
 These include, for example, scenarios with heavy scalars 
\cite{Nima,Guidice,focuspoint,Djouadi2} or some superstring-inspired models 
\cite{u1y}. In addition, one finds studies on the phenomenological
implications of non-universal scalars \cite{ewsbnus,morenus}, 
particularly relating to dark matter \cite{DM1}. The special
thrust of the present work lies in its generality as well
as the emphasis on the relative strengths of different signals
in eliciting a non-universal scalar mass pattern. 

The most important signals of $R$-parity conserving SUSY
consist in large missing transverse energy ${E}_T\!\!\!\!\!\!/~$ ,
 accompanied with energetic jets and leptons of various
multiplicity in the central region. While the signal
strengths, kinematics and event topology of a given
final state yield information of the mass scale of
new particles, it is emphasized that {\em the relative strengths
of different signals corresponding to the spectrum of a given
type often tells us more.} In particular, the departure from the 
mSUGRA scenario can crucially affect some particular
final state. Hence we advocate the detailed exploration of
the `signature space' 
\cite{Subho,Bourjaily:2005ja,CDF,Ash,Barger,Matchev,Baer,ATLAS}
as a whole, and illustrate such exploration
for some representative cases through a multichannel analysis
\cite{baer11}.

Our (restricted) signature space consists of the 
following finals states:
$jets+{E}_T\!\!\!\!\!\!/~$, $same$-$sign$ as well as 
$opposite$-$sign~dileptons$, and  $trileptons$ 
along with $jets+{E}_T\!\!\!\!\!\!/~$. 
In addition, we include the so-called `hadronically
quiet' trilepton events in our analysis. The event rates
predicted are after the imposition of cuts aimed at reducing
the SM backgrounds. We present the ratios of various 
types of final states, thus also reducing uncertainties due
to parton distributions, factorization scales, jet energy 
resolutions etc. These ratios, presented as bar graphs,
demonstrate the departure (or otherwise) from what is
predicted in mSUGRA, for superparticle masses in 
different combinations. They can be supplemented by the absolute
rates, too, for (a) information on the overall SUSY masses,
and (b) cases where the rate of one type of event is either
too small or submerged in backgrounds.

In the mSUGRA models, all  low-scale parameters are derived from
a universal gaugino mass ($M_{1/2}$),  a universal scalar mass ($m_0$),
the trilinear soft SUSY-breaking parameter ($A_0$) and the sign of
the Higgsino mass parameter (sgn($\mu$)) for each value of
$\tan\beta$, the ratio of the two Higgs vacuum expectation 
values (vev)\cite{Martin,mSUGRA}. Since the consequence of gaugino
non-universality \cite{Subho,Ellis,nonugmpheno1,nonugmpheno2,Choi:2007ka,Bt1} 
has been probed in our earlier work, the gauginos have been 
taken to have a universal mass at high scale in this study.

Specifically, we consider three different types of non-universal 
scenarios. These are (a) non-universality of the squark and and
slepton masses, (b) non-universality of the third family
sfermions with respect to the first two, and (c) non-universality
due to high-scale $D$-terms, pertinent to an $SO(10)$ model.
While the first scenario is purely phenomenological, the second
one is motivated by the so-called `inverted hierarchy' at a high
scale, which is advocated
as a solution to the flavour problem \cite{3rd1,3rd2,3rd3,3rd4,3rd5}. 
The third case concerns
a particular theoretical picture where physics between the Planck 
and GUT scales affects the masses of sfermions in different
sub-representations of $SO(10)$, leading to different low-energy
mass patterns \cite{so101,so102,Datta:1999uh}.

The approach advocated here can be useful in so called 
'inverse problem' approach \cite {Nima2}, where one aims to construct 
an underlying theory from a multichannel assortment of data.

The paper is organized as follows. In the following section, we 
outline the general strategies of our collider simulation, including
the main event selection criteria. We discuss the non-universality of
squark-slepton masses and the different predictions in the signature 
space in section 3. Section 4 contains a comparative study of different 
signals at the LHC, when the non-universality is limited to
scalars in the third family. Signatures of
$SO(10)~ D$-terms leading to non-universality is
discussed in section 5, where  we also discuss the 
variation of the mass spectrum with the $D$-term contribution treated as
a free parameter. We summarize and conclude in section 6.
Salient features of the particle spectra in the different cases, 
and the absolute rates of predicted events, are presented
in Appendices A and B, respectively.

\vspace{0.4 cm}

\section {Strategy for simulation}

Before we proceed to analyze specific scenarios,
let us summarize the collider simulation procedure
that has been adopted in all the cases.
The spectrum generated by {\tt SuSpect} v2.3
 \cite{SuSpect} as described in each scenario
is fed into the event generator {\tt Pythia} 6.405 \cite{PYTHIA} 
by {\tt SLHA} interface \cite{sLHA} for the simulation of $pp$ collision 
with centre of mass energy 14 TeV.

We have used {\tt CTEQ5L} \cite{CTEQ} parton distribution functions, 
the QCD renormalization and factorization scales being
both set at the subprocess centre-of-mass energy  $\sqrt{\hat{s}}$.
Other options such as the  scales set at the average mass
of the particles produced in the initial hard scattering
are not found to alter the qualitative features of our
results.  All possible SUSY processes and decay chains consistent 
with conserved $R$-parity have been kept open. 
In the illustrative study presented here, we have switched
off initial and final state radiations. This does
not affect the major conclusions, 
as events with $\ge$2 jets
are mostly considered and jet counting
is not of any crucial significance here.The effect of multiple 
interactions has been neglected. However, we take 
hadronization into account using the fragmentation functions 
inbuilt in {\tt Pythia}.

The final states studied here are \cite{Ash,ATLAS,Baer2}:

\begin{itemize}
 \item Opposite sign dilepton ($OSD$) :
 $(\ell^{\pm}\ell^{\mp})+ (\geq 2)~ jets~ + {E_{T}}\!\!\!\!/$ 
  
\item Same sign dilepton ($SSD$) : 
$(\ell^{\pm}\ell^{\pm})+ (\geq 2)~jets~ + {E_{T}}\!\!\!\!/$

\item Trilepton $(3\ell+jets)$: 
$3\ell~ + (\geq 2) ~jets~ + {E_{T}}\!\!\!\!/$

\item Hadronically quiet trilepton  $(3\ell)$:
$3\ell~ + {E_{T}}\!\!\!\!/$   

\item Inclusive jets ($jets$): $(\geq 3) ~jets~ + X + {E_{T}}\!\!\!\!/$    
\end{itemize}

\noindent
where $\ell$ stands for electrons and or muons.

It should be noted that hadronically quiet trileptons have been
introduced as a separate channel of study here, contrary, for example,
to the one presented in reference \cite{Subho}. 
The reason for our optimism about
this channel is the fact that the very notion of sfermion non-universality
entails scenarios with sleptons that are light with respect to
charginos and neutralinos, a feature that serves to enhance the rates
of final states with high lepton multiplicity arising from decays
of the latter. The numerical results presented in the following sections
show that, with exceptions, this optimism is not entirely misplaced.

We have generated all dominant SM events  
in {\tt Pythia} for the same final states, using the same  
factorization scale, parton distributions and cuts. 
$t\bar t$ production gives the most serious backgrounds 
in all channels excepting in the trilepton channels,
for which electroweak backgrounds can be serious.
For the inclusive jet signals, the final states without any 
isolated, central, hard leptons are also prone to  
large QCD backgrounds, where, for example, jet energy
mismeasurement can lead to a tail with missing-$E_T$.
The maximum reduction of such QCD backgrounds
is very challenging (especially
due to uncertainties in the prediction and interpretation
of multi jets). In our theoretical study, keeping
the above problem in mind, we have
tried to be conservative by imposing a cut of
100 GeV on {\em each jet} and not choosing to order
their hardness cuts. While one can further improve
on this by making the ${E_{T}}\!\!\!\!/$ cut even higher,
our  main message, namely, the sensitivity of the ratios
of various signals to different non-universal scenarios,
still retains its relevance after such improvements.

The signal and background events have been all calculated for
an integrated luminosity of 300 fb$^{-1}$. 
As noted earlier, the  event ratios 
which are the primary objects of our analysis
help in avoiding  uncertainties in prediction.
Cases where the number of signal events in any of the channels
used in the ratio(s) is less than three have been left out.
Also, in the histograms (to be discussed in the next section),
cases where any of the entries in the ratio has
a significance less than 2$\sigma$
have been specially marked with a {\bf \#} in the bar graphs.
since our observations on them may still be useful 
if statistics can be improved.

The cuts used in our analysis are as follows:

\begin{itemize}
\item  Missing transverse energy $E_{T}\!\!\!\!/$ $\geq ~100$ GeV.

\item ${p_{T}}^l ~\ge ~20$ GeV and $|{{\eta}}_{\ell}| ~\le ~2.5$.
 
\item An isolated lepton should have lepton-lepton separation
 ${\bigtriangleup R}_{\ell\ell}~ \geq 0.2$, lepton-jet separation 
 ${\bigtriangleup R}_{{\ell}j}~ \geq 0.4$, the energy deposit 
due to jet activity around a lepton ${E_{T}}$ within 
$\bigtriangleup R~ \leq 0.2$ of the lepton axis should be $\leq 10$ GeV.

\item ${E_{T}}^{jet} ~\geq ~100$ GeV and $|{\eta}_{jet}| ~\le ~2.5$

\item For the hadronically quiet trilepton events, we have used 
in addition, invariant mass cut on the same flavour opposite
sign lepton pair as $|M_{Z}-M_{l_{+}l_{-}}| ~\geq 10$ GeV. 

\end {itemize}

\noindent
where  $\bigtriangleup R = \sqrt {{\bigtriangleup \eta}^2
+ {\bigtriangleup \phi}^2}$ 
is the separation in pseudo rapidity and azimuthal angle plane. 

Jets are formed in {\tt Pythia} using {\tt PYCELL} jet formation criteria
with $|{\eta}_{jet}| ~\le ~5$ in the calorimeter, 
$N_{\eta_{bin}}=100$ and $N_{\phi_{bin}}=64$. 
For a partonic jet to be considered as a jet initiator $E_{T}> 2$ GeV 
is required while a cluster of partonic jets to be called a hadron-jet
 $\sum_{parton} {E_{T}}^{jet}$ is required to be more than 20 GeV.
For a formed jet the maximum $\bigtriangleup R$ from the jet initiator 
is 0.4.

We have checked the hard scattering cross-sections 
of various production processes with {\tt CalcHEP} \cite{CalcHEP}.
All the final states with jets at the parton level 
have been checked against the results available in \cite{Ash}. 
The calculation of hadronically quiet trilepton 
rates have been checked against \cite{hq3l}, in the appropriate limits.

\section{Squark-slepton Non-universality}

Here we select a scenario where the squarks and slepton 
masses at low-energy are results of evolution from
mutually uncorrelated mass parameters 
($m_{0\tilde{q}}$ and $m_{0{\tilde l}}$ respectively) at a high scale.
Although this is a purely phenomenological approach,
it is helpful in the sense that it embodies the complete
independence of the coloured and uncloured 
scalar masses at the high scale,
while still achieving some simplification of the parameter
space, by avoiding a random proliferation of low-energy 
masses. The choice of parameters made in this manner 
takes all collider and low-energy constraints  into
account, as summarized in the subsection below.

\subsection {Choice of SUSY parameters}

As has been already indicated, we have confined 
ourselves to $R$-parity conserving supersymmetry 
where the lightest neutralino is the LSP.
The squark-slepton spectrum is generated by {\tt SuSpect} v2.3
 \cite{SuSpect} 
with the {\bf pMSSM} option, where  a separate mass
parameter  for squarks and sleptons is assumed at the high
scale. The Higgs mass parameters $m_{H_{u}}^2$ $\&$ $m_{H_{d}}^2$ 
are also taken to evolve from the high-scale slepton mass. 
We tune the non-universal scalar masses
and gaugino masses at the GUT scale such that 
the following combinations arise:\\
$(m_{\tilde g},m_{\tilde q^{1,2} })$ = (500,500), (500,1000) and 
(1000,1000) where $m_{\tilde g}$ is the gluino mass and 
$m_{\tilde q^{1,2} }$ denote the (approximately degenerate) squark masses
of the first two families at the electroweak symmetry breaking (EWSB) scale
defined by the `default option' in {\tt SuSpect}, i.e. 
$\sqrt{m_{\tilde{t_{L}}}m_{\tilde{t_{R}}}}$. All the above masses are in GeV.
All the aforementioned sets are studied for three non-universal 
slepton masses of the first two families $m_{\tilde l^{1,2}}$ 
(approximately degenerate) at the low-scale, 
namely, 250 GeV, 500 GeV and 750 GeV 
with $\tan {\beta}$= 5 and 40 for each choice. 
The high-scale value of the soft trilinear parameter ($A_0$)
has been set at zero, a practice that has been followed 
in the subsequent sections, too (For details see table A1 and A2).  

Radiative electroweak
symmetry breaking has been ensured in each case, after which the
positive value of $\mu$ has been chosen for illustration, in this section
as well as in the subsequent ones.
One also achieves gauge coupling unification 
at high scale and consistency with laboratory constraints 
on a SUSY scenario. Consistency with low-energy FCNC constraints 
such as those from $b\rightarrow s\gamma$, and also with the
data on muon anomalous magnetic moment 
are checked for every combination of parameters \cite{constraints,LEP}
used in the analysis. No constraints from dark matter have been included here.
We have used the strong coupling ${\alpha_3 (M_{Z})}^{\overline{MS}}= 0.1172$ 
for this calculation which is again the default option in {\tt SuSpect}. 
Throughout the analysis we have assumed the top quark mass to be 171.4 GeV.
No tachyonic modes for sfermions are allowed at any energy scale. 
Gaugino masses have been treated as universal at high scale for simplification.

In this study the low energy sfermion masses vis-a-vis those of charginos
and neutralinos primarily dictate the phenomenology. Relating them to 
high scale parameters is done for the purpose 
more in the way of illustration, and achieving a very 
high degree of precision in the relationship among low and high scale 
parameters is not of primary importance here. Thus, in the running of 
parameters, one-loop renormalization group
equations (RGE) have been used.  
No low-energy radiative corrections to the chargino and neutralino masses
matrices have been taken, which
does not affect our analysis in any significant way \cite{Rammond}.
Full one-loop and the dominant two-loop corrections 
to the Higgs masses are incorporated.

\begin{figure}[t]
\begin{center}
\centerline{\epsfig{file=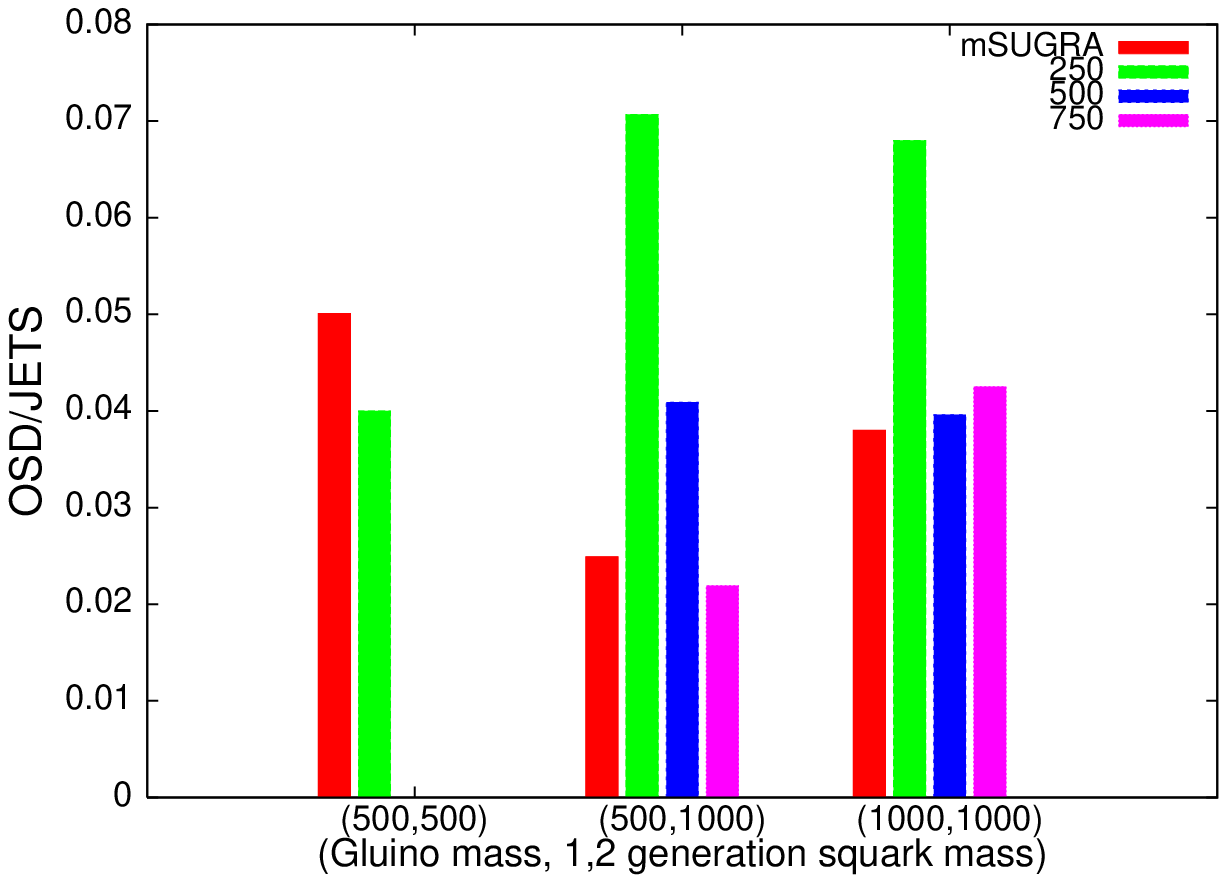,width=6.5 cm,height=5.50cm,angle=-0}
\hskip 20pt \epsfig{file=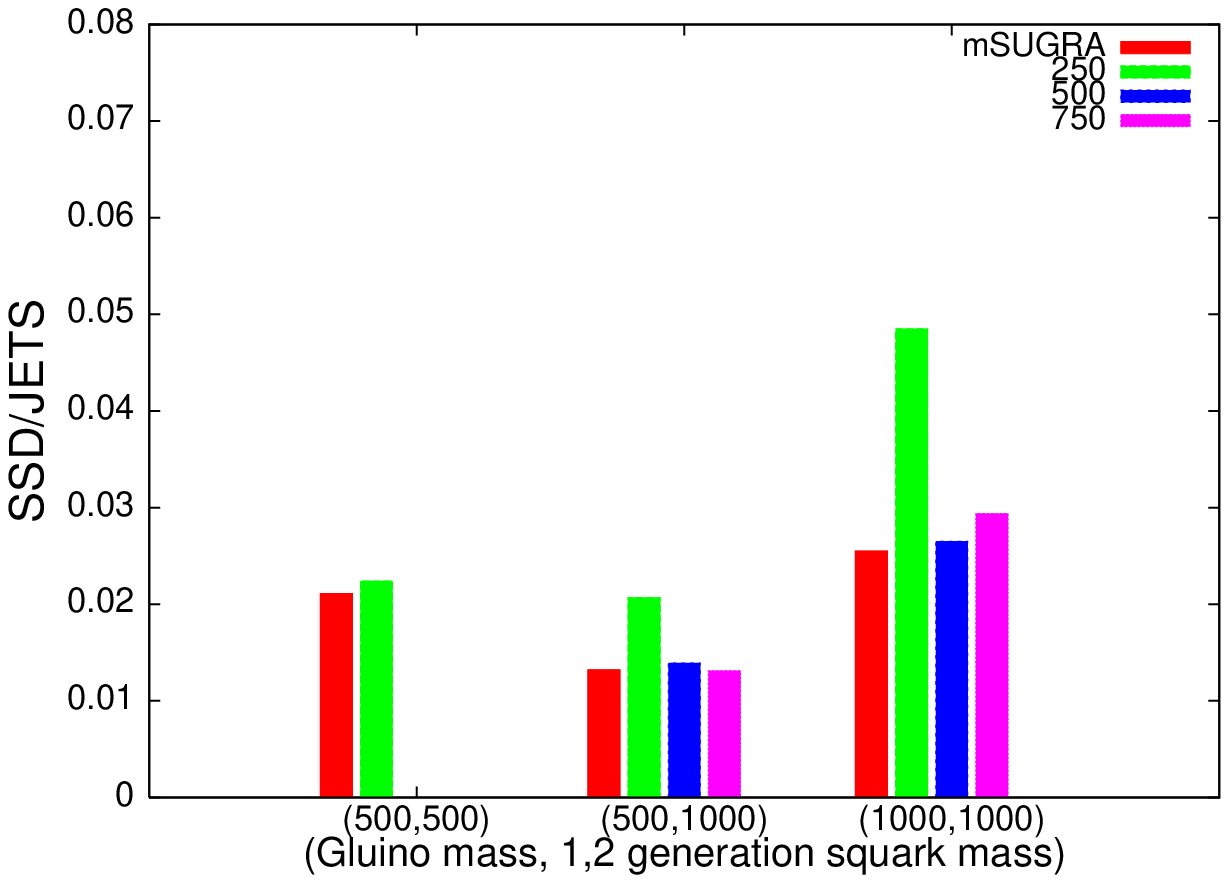,width=6.5cm,height=5.50cm,angle=-0}}
\vskip 10pt
\centerline{\epsfig{file=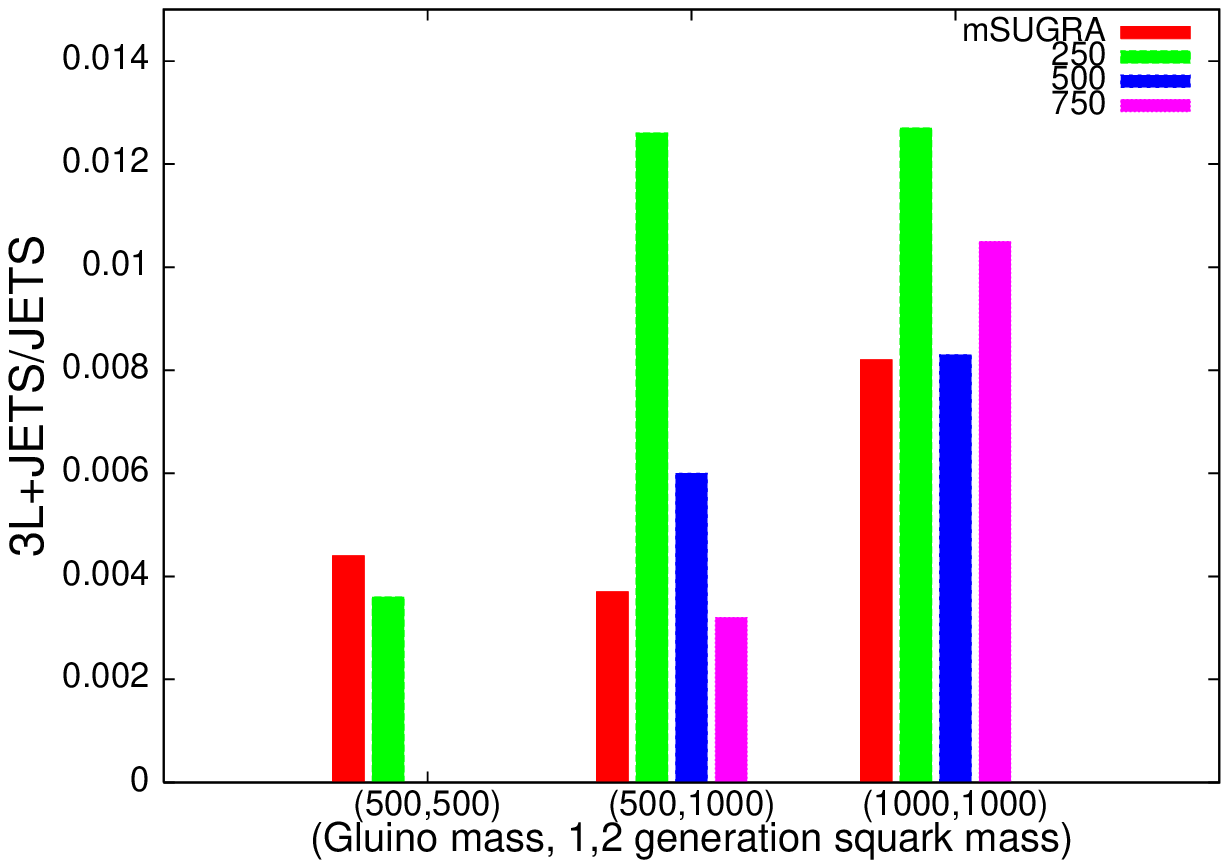,width=6.5 cm,height=5.50cm,angle=-0}
\hskip 20pt \epsfig{file=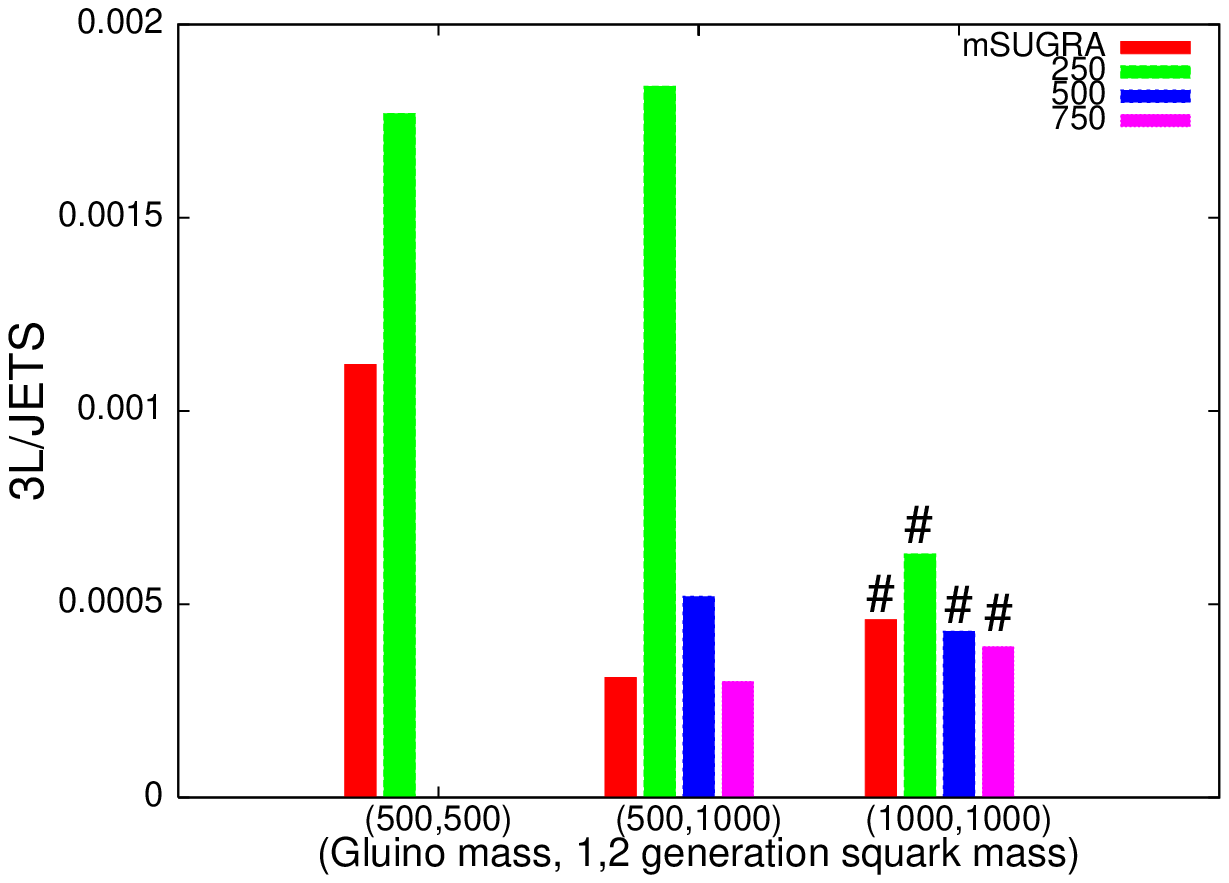,width=6.5cm,height=5.50cm,angle=-0}}
\caption{ Event ratios for Squark-Slepton Non-universality: $\tan{\beta}=5$} 
\end{center}

\end{figure}

\begin{figure}[t]
\begin{center}
\centerline{\epsfig{file=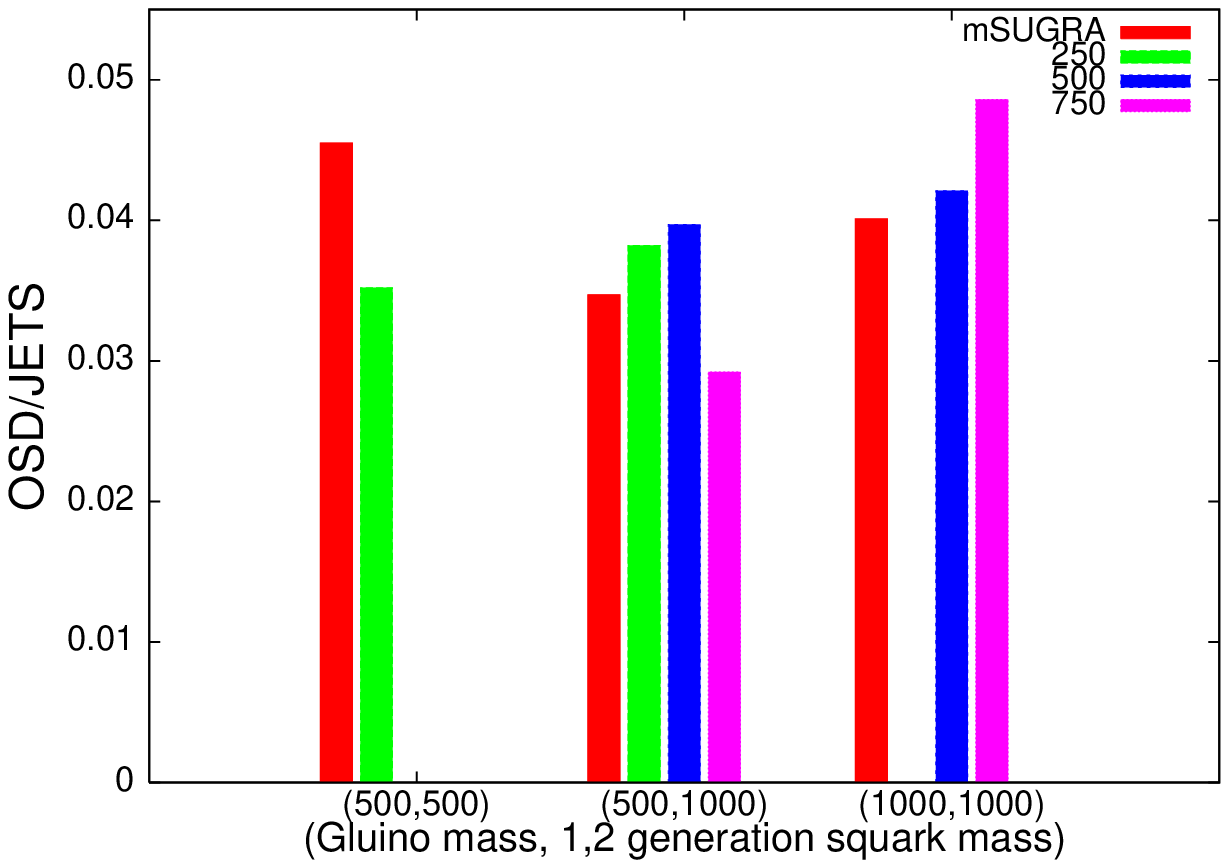,width=6.5 cm,height=5.50cm,angle=-0}
\hskip 20pt \epsfig{file=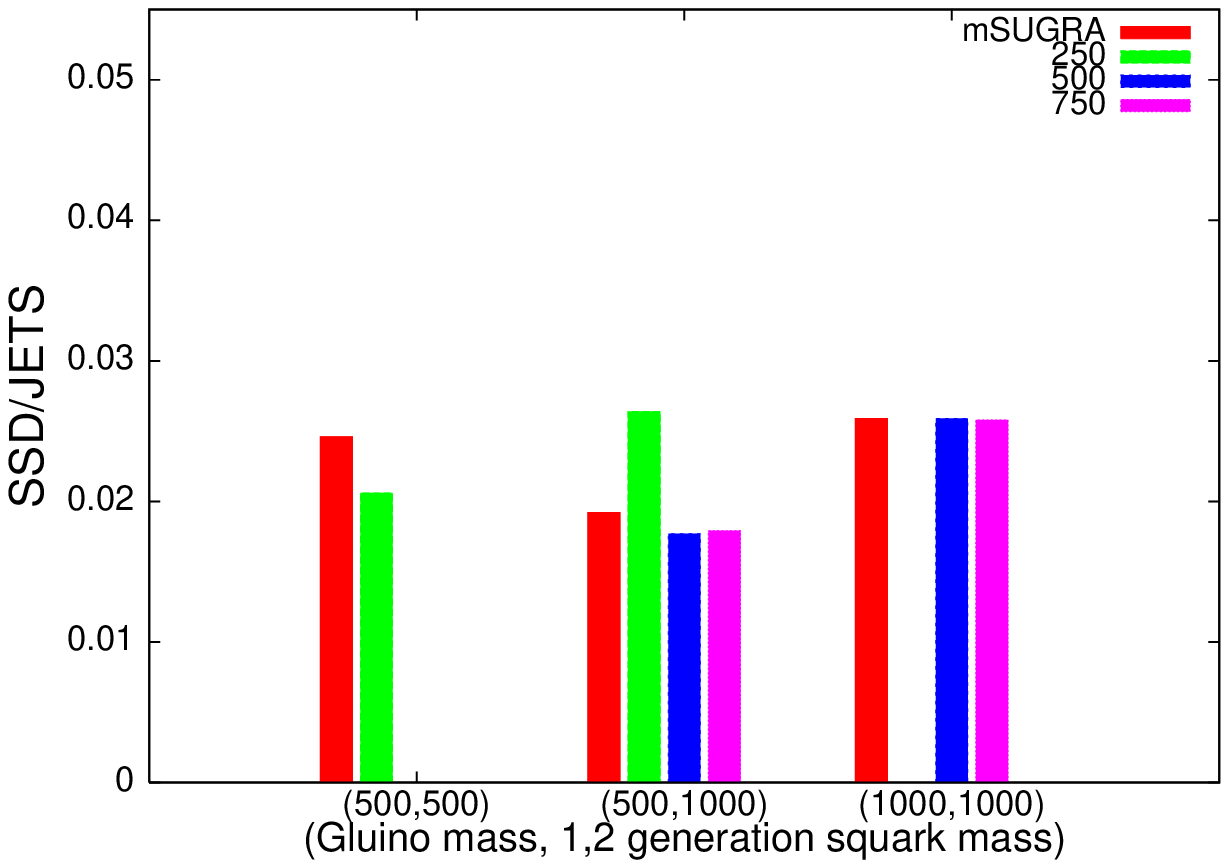,width=6.5cm,height=5.50cm,angle=-0}}
\vskip 10pt
\centerline{\epsfig{file=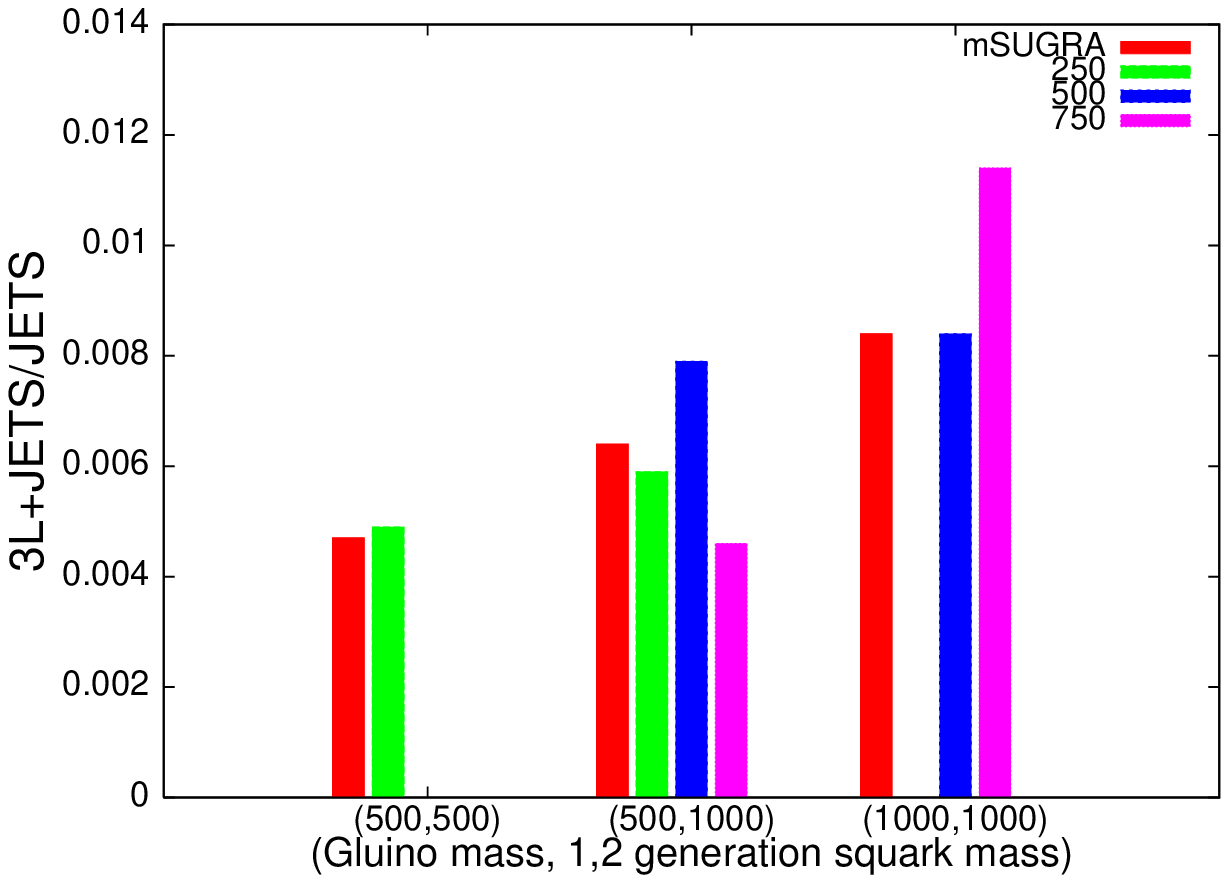,width=6.5 cm,height=5.50cm,angle=-0}
\hskip 20pt \epsfig{file=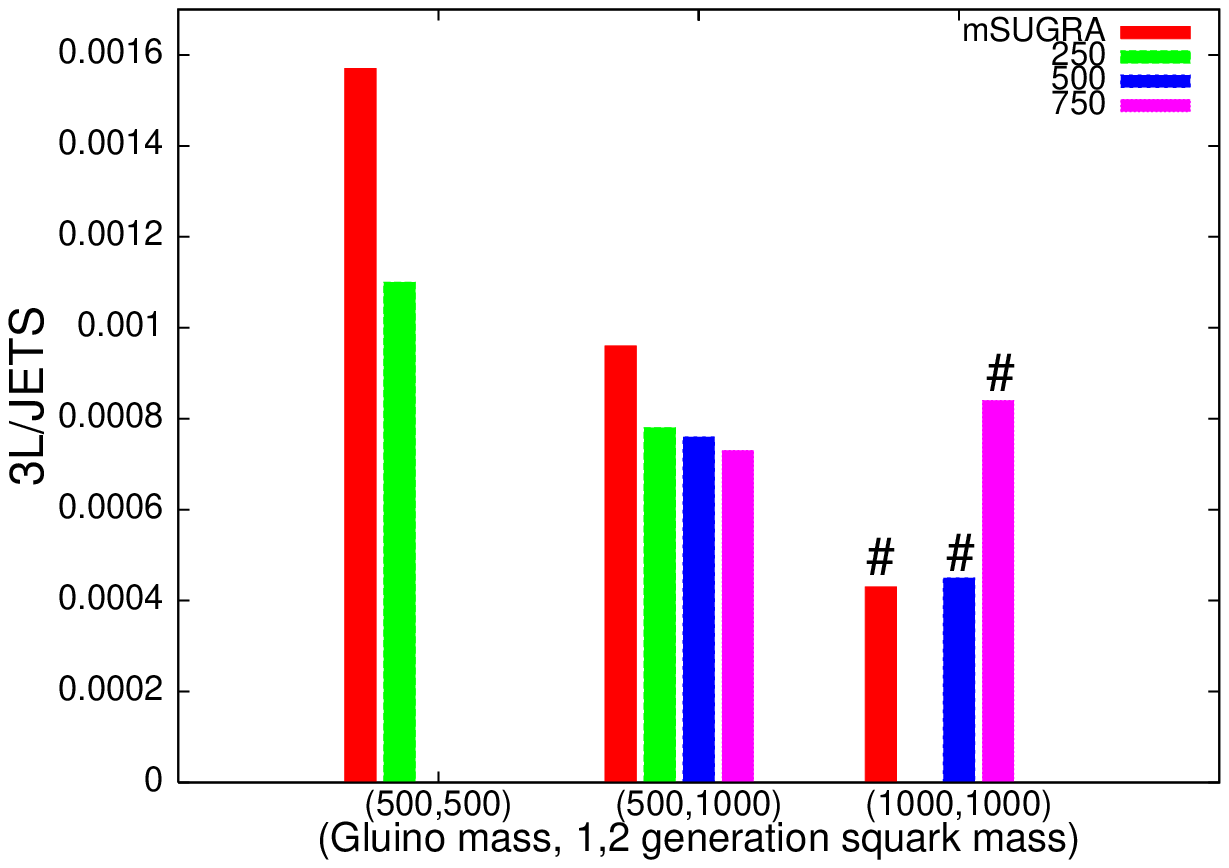,width=6.5cm,height=5.50cm,angle=-0}}
\caption{ Event ratios for Squark-Slepton Non-universality: $\tan{\beta}=40$} 
\end{center}

\end{figure}

\subsection{Numerical results}   

In figures 1 and 2, we have presented four ratios, namely, 
$OSD/jets$, $SSD/jets$, $(3\ell+jets)/jets$ and $3\ell/jets$.
For $(m_{\tilde g},m_{\tilde q^{1,2} })$ = (500,500) GeV, electroweak
symmetry breaking conditions are not satisfied when the low-energy 
slepton mass is 500 or 750 Gev. This is because, with gaugino masses 
on the lower side, such large slepton masses require a rather large
high-scale value for the slepton-Higgs mass parameter, which
prevents  $m_{H_{u}}^2$ from being driven to a negative value at
the electroweak scale. For low slepton and high gaugino masses,
on the other hand, the lighter stau eigenstate becomes the LSP
for $\tan\beta$ = 40.

A survey of figures 1 and 2 reveals the following general 
features for the case with squark-slepton non-universality:

\begin{itemize}
\item The case with the lowest choice of slepton masses,
namely, $m_{\tilde l^{1,2}} = 250$ GeV, is fairly
distinguishable from the others, especially for the 
squark masses on the higher side. This is primarily because
low-lying sleptons participate in the chargino and
neutralino cascades to yield more events with leptons
in the final state. Such an effect is noticeable
for $\tan \beta =$ 5. One has to remember here that
the chargino and neutralino mass matrices whose textures
govern the cascades are also controlled by $\mu$  which
is related to $m_{\tilde l^{1,2}}$. Thus the final rates
depend on a crucial interplay of the slepton mass parameter,
the gaugino masses and $\tan \beta$, over and above the
enhanced probability of on-shell decays of charginos
and neutralinos into sleptons.

\item Cases with  $m_{\tilde l^{1,2}} = 500$ GeV are
by and large difficult to differentiate from a spectrum
with universal scalar mass.

\item  The $3\ell+jets$ events allow one 
to distinguish cases with the slepton mass on the high 
side, such as $750$ GeV. This effect is more prominent
for high gluino mass and large $\tan \beta$.

\item The hadronically quiet trilepton signals give us 
sufficient distinction in cases where the background is
not forbidding. This channel gets drowned in backgrounds
only for $(m_{\tilde g},m_{\tilde q^{1,2} })$ = (1000,1000) GeV. 
The universal case is best distinguished with one where the slepton 
mass of the first two families assumes the lowest chosen value 
(250 GeV). This is because these would help on-shell 
slepton production in two-body
decays of charginos and neutralinos. Naturally, higher gluino masses
hurt this channel because they mean higher chargino/neutralino
masses and thus lower production rates with gaugino universality 
(see table B1 and B2). 
Moreover, the distinction is more prominent
for $\tan \beta$ on the lower side.

\item In general (including the difficult case mentioned
above), trileptons in the final state are  the most
useful signals in distinguishing among different scenarios. 
\end{itemize}

\section{Non-universality in the third family}

In order to address the FCNC problem that continues to haunt
SUGRA-type models, it has sometimes been proposed that
the first two families of squarks and sleptons are very heavy.
This suppresses FCNC in most cases. At the same time, a third
family of sfermions within a TeV suffices to provide a solution 
to the naturalness problem. Such scenarios have been theoretically
motivated, for example, in string-inspired models, assuming 
flavour-dependent coupling to modular fields, or postulating that
the masses of the third family scalars arise from a separate
$F$-term vev. $D$-terms of an anomalous $U(1)$ symmetry have also been
suggested for implementing such `inverted hierarchy'
\cite{3rd1,3rd2,3rd3,3rd4,3rd5}. 

Since this is a rather representative case of scalar
non-universality, we have subjected the resulting  spectra
to the multichannel analysis outlined earlier. However, 
we do not confine ourselves to any special theoretical
scenario, except assuming that scalar masses in the third 
family evolve from a separate high-scale mass parameter $m^3_0$, 
while a different parameter $m^{(1,2)}_0$ is the origin of scalar
masses in the first two families.

\subsection{Choice of parameters}

As has been already mentioned, we have assumed the third family
scalar masses to arise out of a separate parameter at high scale ($m^3_{0}$).
The SUSY breaking mass parameters $m_{H_u}~ \& ~m_{H_d}$ in the Higgs
sector are also assumed to originate in same parameter $m^3_{0}$.
Otherwise, in cases where  $m^{(1,2)}_0$ is very high and essentially
decoupled, a correspondingly high value of the Higgs mass parameter(s)
will make it difficult to obtain electroweak symmetry breaking in 
a consistent manner.

This allows one to fix the magnitude of the $\mu$-parameter,
which we have taken to be of positive sign throughout 
our analysis. As in the previous section, we have taken
$A_0 =$ 0.
The unification of gaugino masses and gauge couplings at
high scale has been ensured. 
As before, the  {\bf pMSSM} option in {\tt SuSpect}  has been used,
and $m^3_0$ as well as the high-scale gaugino mass 
parameter has been tuned in such a 
way as to yield specific values of the gluino mass and the lighter 
stop mass ($m_{\tilde{t_1}}$) at low-energy. The chosen combinations 
of $(m_{\tilde g},m_{\tilde{t_1}})$ are (500,500), (500, 1000)
and (1000,1000), all masses being expressed in GeV.
These values are used in the labels of the x-axis in figures 3 and 4.

For each combination mentioned above, two
choices of $m^{(1,2)}_0$ have been made, corresponding
to the average squark mass in the first two families equal 
to 1 TeV and 10 TeV, respectively, at the electroweak scale.
It should be mentioned here that a parameter combination
with $m^{(1,2)}_0$ of the order of a few TeV's and
the third family squark masses around a few hundred GeV's 
is admissible even in an mSUGRA scenario, where the
first two families of squarks can be missed at the LHC \cite{baertata}.   
The results for such choices are juxtaposed with the 
universal SUGRA scenario tuned in such a way as to yield
the same $(m_{\tilde g},m_{\tilde{t_1}})$, 
in the bar graphs shown in figures 3 and 4. 
Two values of $\tan \beta$, namely, 5 and 40, have been
used for every combination of masses (see table A3 and A4).

The procedure adopted in running the parameters is
the same as that described in the previous section. 
All constraints on the low-energy parameters, 
including those from FCNC, have been satisfied 
in each case.

\subsection{Numerical results}

The general format of presentation of the numerical results
in this case is similar to that adopted in the previous
section. All the parameter combinations here are found to 
lead to consistent sparticle spectra, satisfying
the electroweak symmetry breaking conditions and other
necessary requirements. 

The various event rates are influenced by some salient 
features of the spectrum. First of all, the high  value of
$m^{(1,2)}_0$ required to make the first two squark 
families as heavy as 10 TeV leaves little significance for
gaugino corrections at low scale, resulting in the 
close degeneracy of squark and sleptons in the first 
two families. For the squark masses around 1 TeV for the
first two families, on the other hand, one has to take a 
much smaller $m^{(1,2)}_0$, which leads to relatively
light sleptons. For the third family, the effects
of mixing and Yukawa coupling bring the lighter stop
below all other sfermions, the difference being
more pronounced for low $\tan \beta$ (see table in B3 and B4).
 
\begin{figure}[t]
\begin{center}
\centerline{\epsfig{file=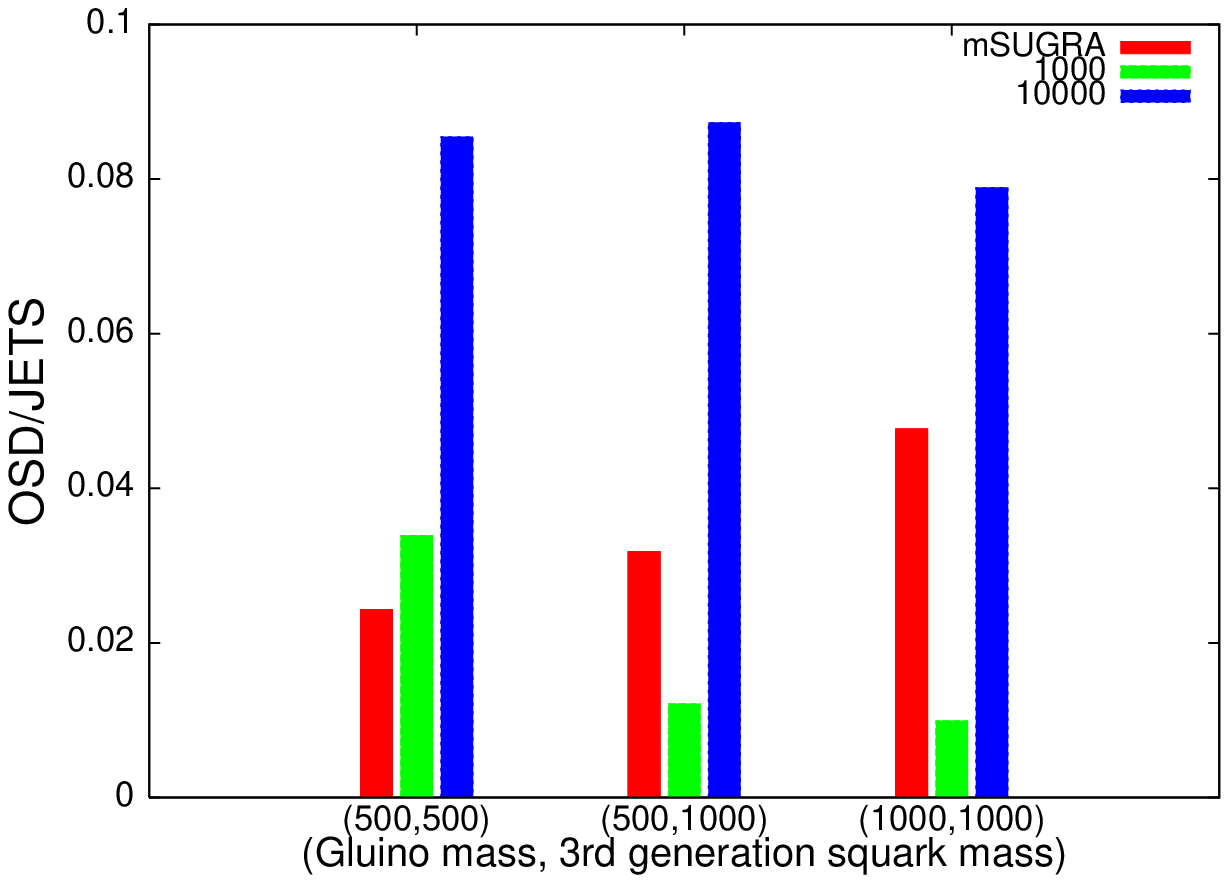,width=6.5 cm,height=5.50cm,angle=-0}
\hskip 20pt \epsfig{file=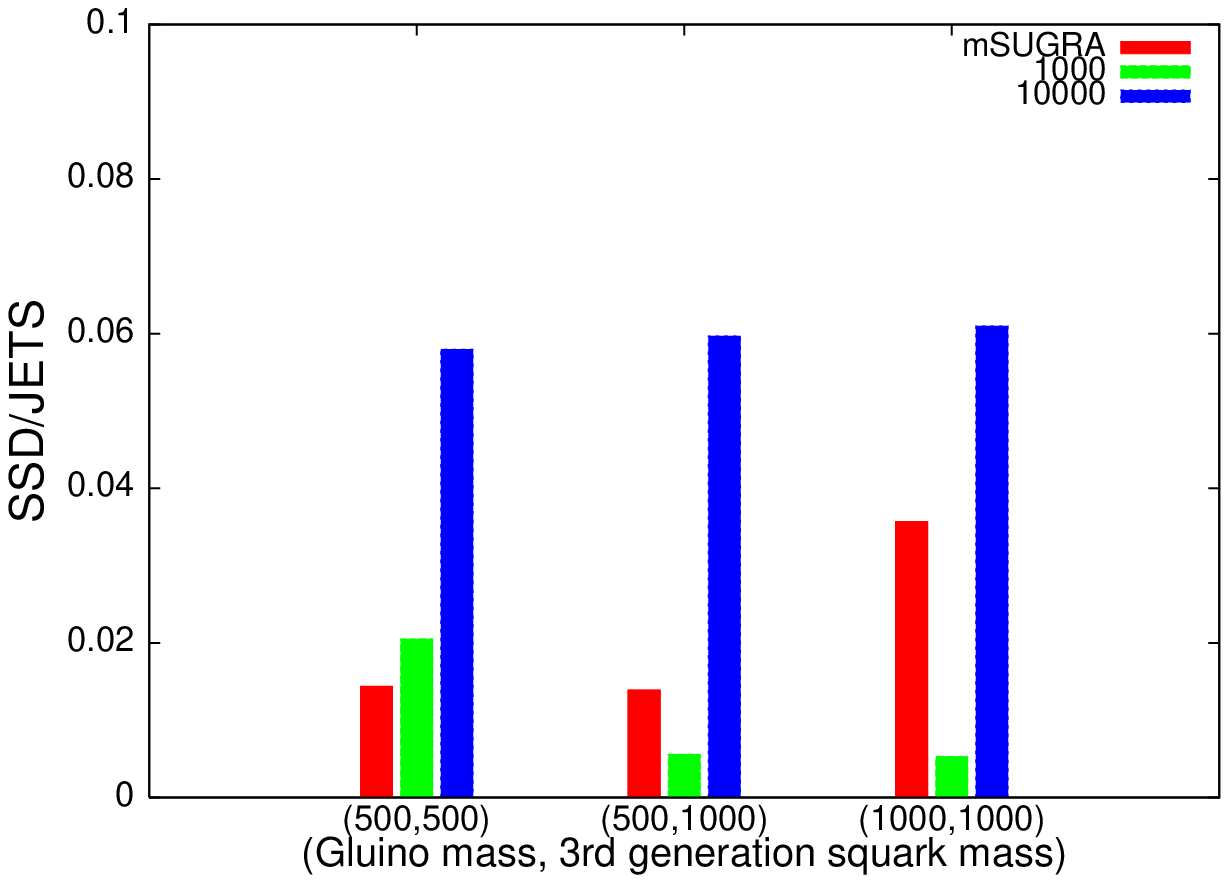,width=6.5cm,height=5.50cm,angle=-0}}
\vskip 10pt
\centerline{\epsfig{file=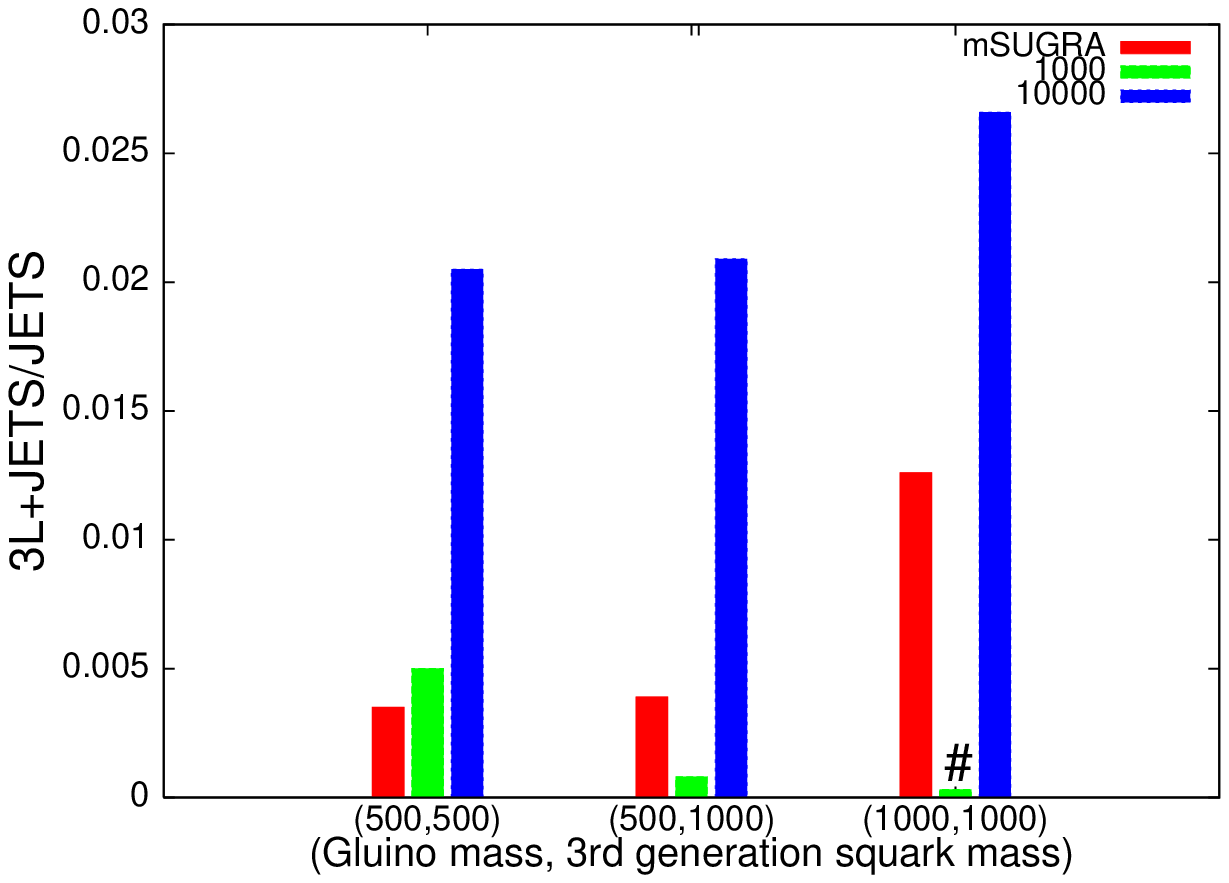,width=6.5 cm,height=5.50cm,angle=-0}
\hskip 20pt \epsfig{file=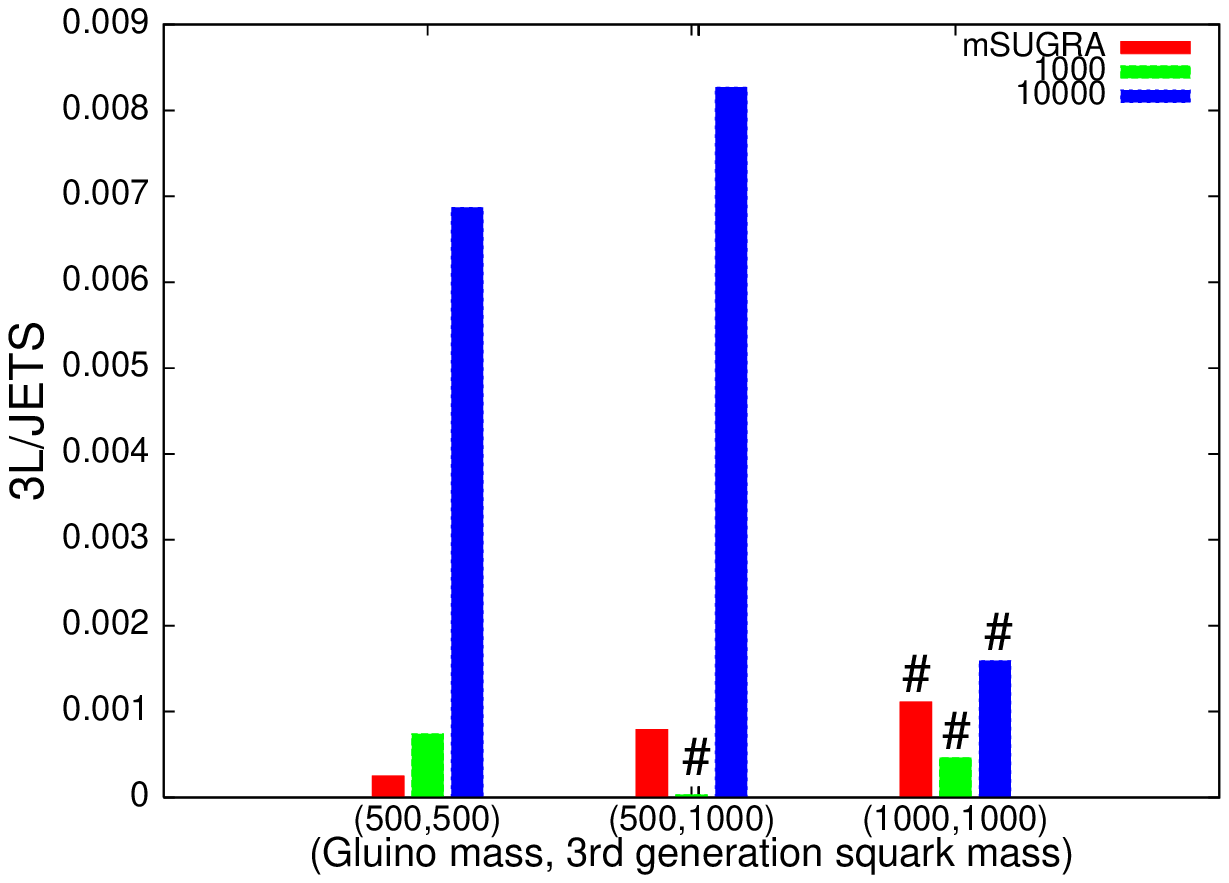,width=6.5cm,height=5.50cm,angle=-0}}
\caption{ Event ratios for 3rd family scalar Non-universality: 
$\tan{\beta}=5$} 
\end{center}

\end{figure}

\begin{figure}[t]
\begin{center}
\centerline{\epsfig{file=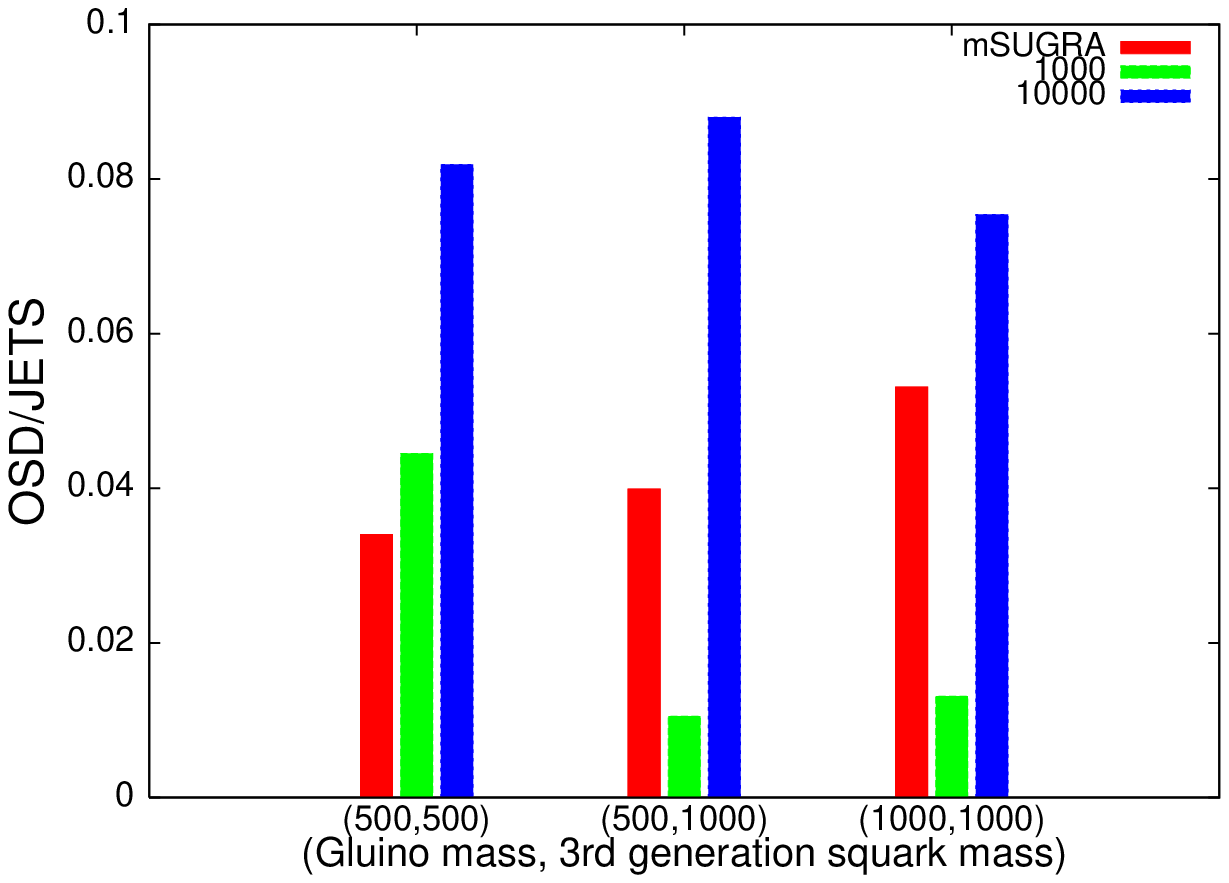,width=6.5 cm,height=5.50cm,angle=-0}
\hskip 20pt \epsfig{file=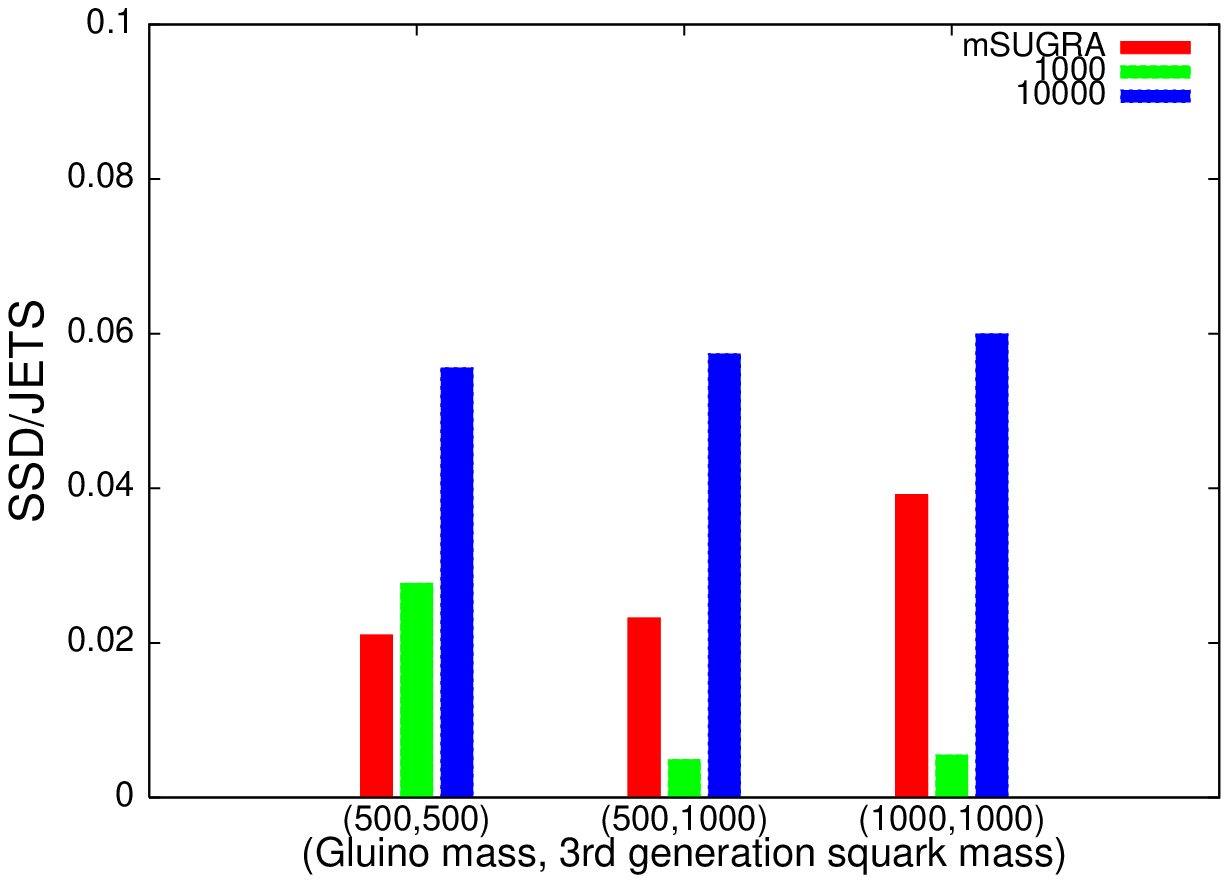,width=6.5cm,height=5.50cm,angle=-0}}
\vskip 10pt
\centerline{\epsfig{file=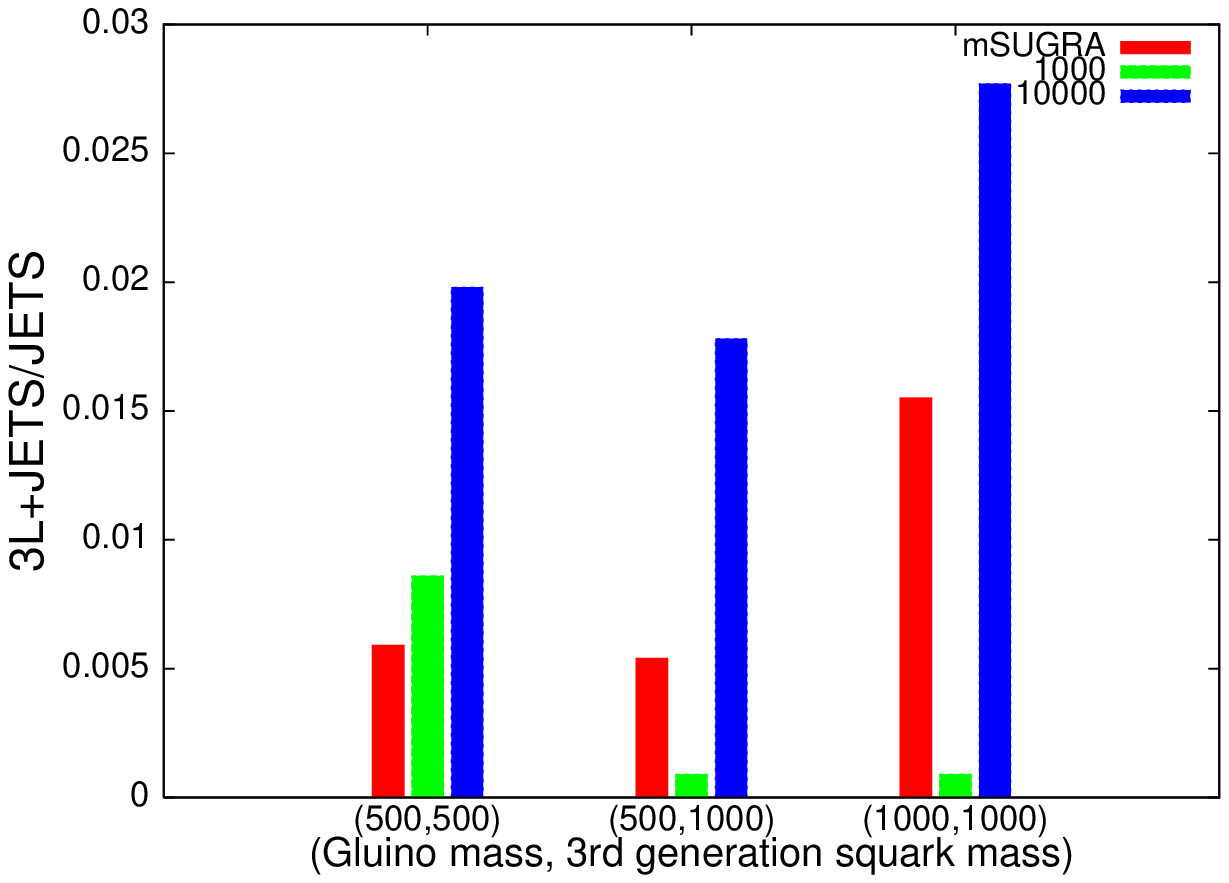,width=6.5 cm,height=5.50cm,angle=-0}
\hskip 20pt \epsfig{file=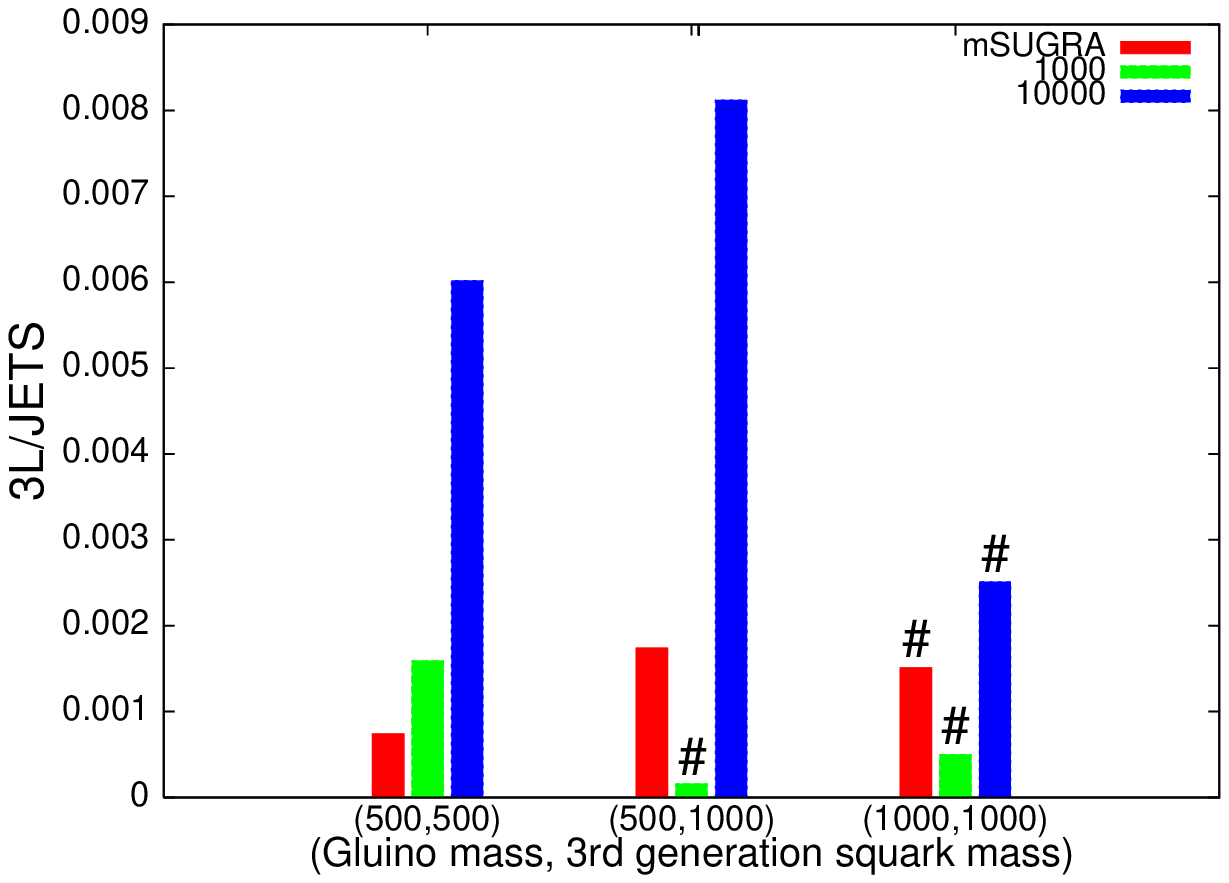,width=6.5cm,height=5.50cm,angle=-0}}
\caption{ Event ratios for 3rd family scalar Non-universality:
$\tan{\beta}=40$} 
\end{center}

\end{figure}

The following broad features can be seen in the results:

\begin{itemize}
\item The rate of leptonic events relative to the
all-jet final state goes up significantly for higher
masses for the first two generations, i.e. 
$m_{\tilde{q}^{1,2}} = 10$ TeV. This is because final states in the
cases of decoupled first two families are dominated by the
stop, which leads to more avenues of lepton production
via top decay. The relative suppression of all-jet 
events from squark pairs (of the first two families) 
is also responsible for lower values of the denominators
in different ratios.

\item The leptonic final states for the non-universal 
case with  $m_{\tilde{q}^{1,2}} = 1$ TeV get considerably depleted
with respect to the corresponding universal cases, especially
for relatively high third family squark masses. 
This happens as a result of our parametrization
where we are matching the mass of the lighter stop
between the two cases. While this means
heavier squarks of the first two families in universal case, 
the non-universal case with $m_{\tilde{q}^{1,2}} = 1$ TeV gives such 
squarks in the same mass range. Therefore, they contribute
more effectively to all-jet final states, leading to a
depletion of leptonic signals.
This feature is reflected not only in the various ratios
but also in the absolute values of the events rates.

\item The difference between 
$m_{\tilde{q}^{1,2}} = 1$ TeV $m_{\tilde{q}^{1,2}} = 10$ TeV 
is most clearly noticeable for the trilepton channel.

\item In a way similar to the other ratios, higher values 
of third family scalar masses facilitate
distinction via the hadronically quiet channels. However, this
channel does not really serve as a better discriminator than 
$OSD$, $SSD$, and inclusive trilepton final states for this
type of non-universality. The underlying reason for this
is again the enhancement of the latter through frequent occurrence
of the top quark in SUSY cascades. Also, just as for
squark-slepton non-universality, the hadronically quiet trileptons
are suppressed by backgrounds for 
$(m_{\tilde g},m_{\tilde{t_1}})$ = (1000,1000) GeV.

\item Unlike the other cases of non-universality studied in this paper
 the observed features bear very little imprint on the value of $\tan \beta$.

\end{itemize}

\section{Non-universality due to $SO(10) ~D$-terms}

In the two previous sections,  scenarios reflecting 
scalar non-universality have been considered in a purely
phenomenological ways. Now we take up a specific theoretical
model, namely one based on an $SO(10)$ SUSY Grand Unified 
Theory (GUT) \cite{GUT}.

In an $SO(10)$ framework, the matter fields belong to
the representation {\bf 16}, and can be further classified
into sub-multiplets, depending on the representations of
$SU(5)$ to which they belong. In this classification, 
expressing the (s)fermions generically to include all families,  
the superfields $D^c$ and $L$  belong to $\bar{\bf 5}$, while
$Q$, $U^c$ and $E^c$ belong to {\bf 10}, where $Q$ and $L$ denote
$SU(2)$ doublets and the others, singlets. The breakdown of $SO(10)$
(without any intermediate scale) to the SM gauge group, which
amounts to a reduction of rank, will therefore
endow the scalars in these different $SU(5)$ representations
with different $D$-terms \cite{so101}. Consequently, the high-scale scalar 
mass parameters will be different for the two multiplets respectively for
$\bar{\bf 5}$ and {\bf 10}: \cite{so102,Datta:1999uh}

\begin{eqnarray}
m^2_{\bar 5} = m^2_0 - 1.5 D m^2_0 ~~~ (for~ D^c~ \&~ L) \\
m^2_{10} = m^2_0 + 0.5 D m^2_0  ~~~(for~ E^c, U^c~  \&~ Q)\\
\end{eqnarray}
\noindent
thus leading to a predestined non-universality in the GUT scale
itself. Here D is a dimensionless parameter quantifying the
added contribution to the SUSY breaking terms in terms of 
the `universal' high-scale mass parameter $m_0$.

\subsection{Choice of parameters}

We have restricted the value of $D$ in order to avoid 
tachyonic modes at high scale. Thus
$D$ = 0.5, -0.5 and -1.25 have been taken, $m_0$ being fixed at
300 GeV. $M_{1/2}$ has been chosen in such a way as to obtain the
low-scale gluino mass at 500 GeV, 1 TeV and 1.5 TeV.

While the sign of $\mu$ has been kept positive in each case,
we have again chosen $\tan\beta$ to be 5 and 40. The low-energy
spectrum is the result of one-loop RGE following {\tt Suspect},
with the {\bf pMSSM} option (see table A5 and A6).

\begin{figure}[t]
\begin{center}
\centerline{\epsfig{file=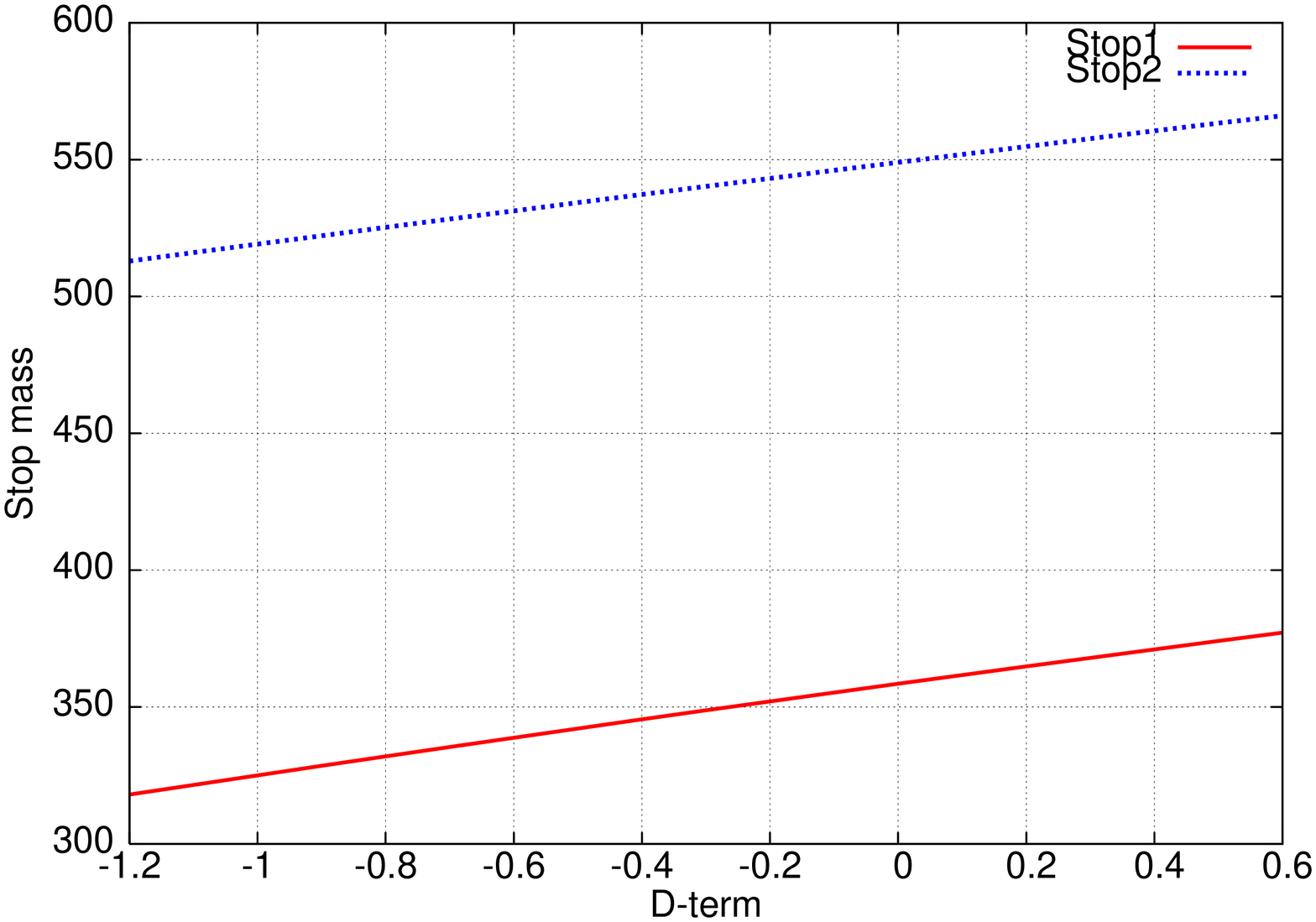,width=5.5 cm,height=6.50cm,angle=-90}
\hskip 20pt \epsfig{file=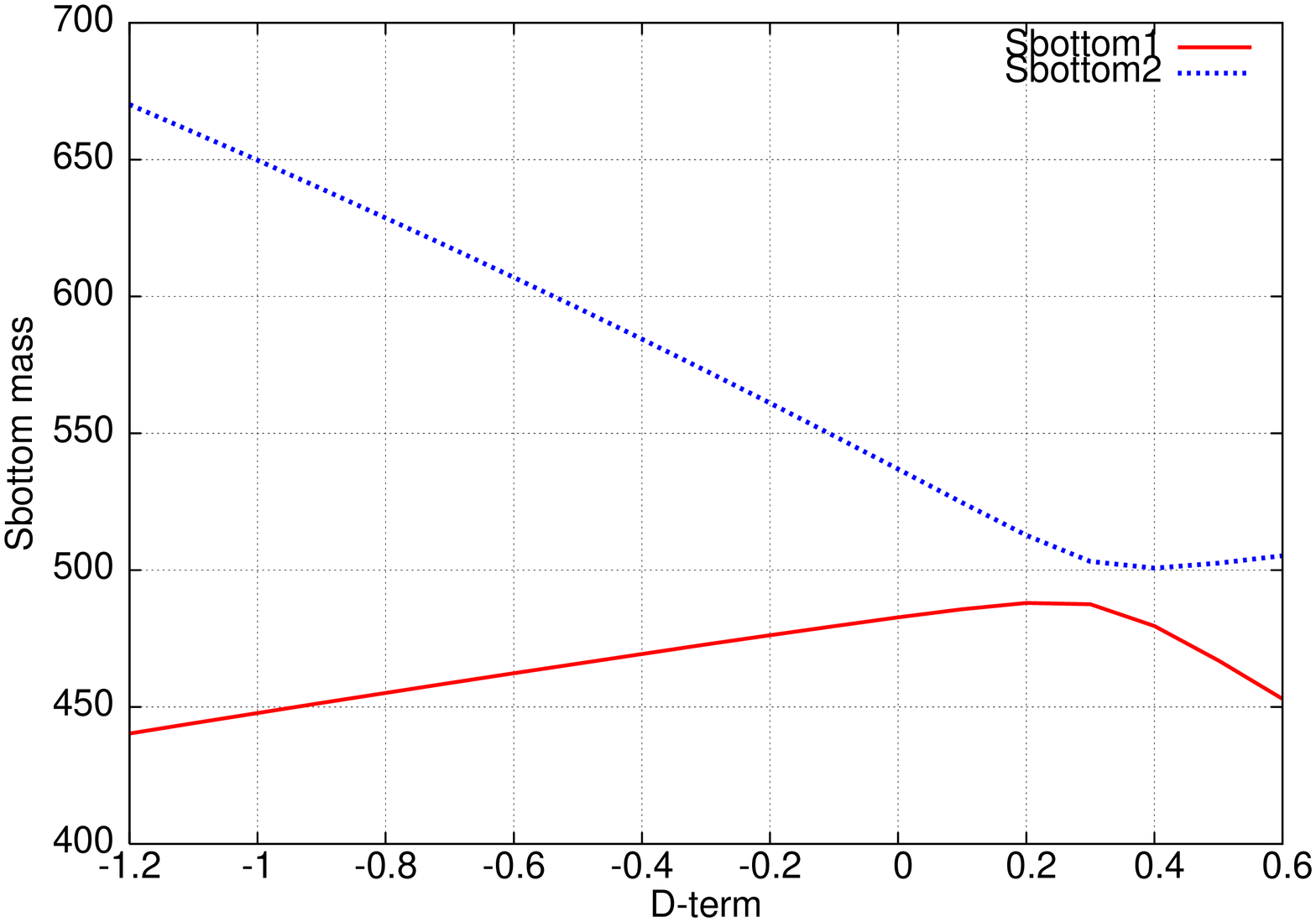,width=5.5cm,height=6.50cm,angle=-90}}
\vskip 10pt
\caption{Variation of stop and sbottom mass with $D$-term:$\tan{\beta}=5$}
\centerline{\epsfig{file=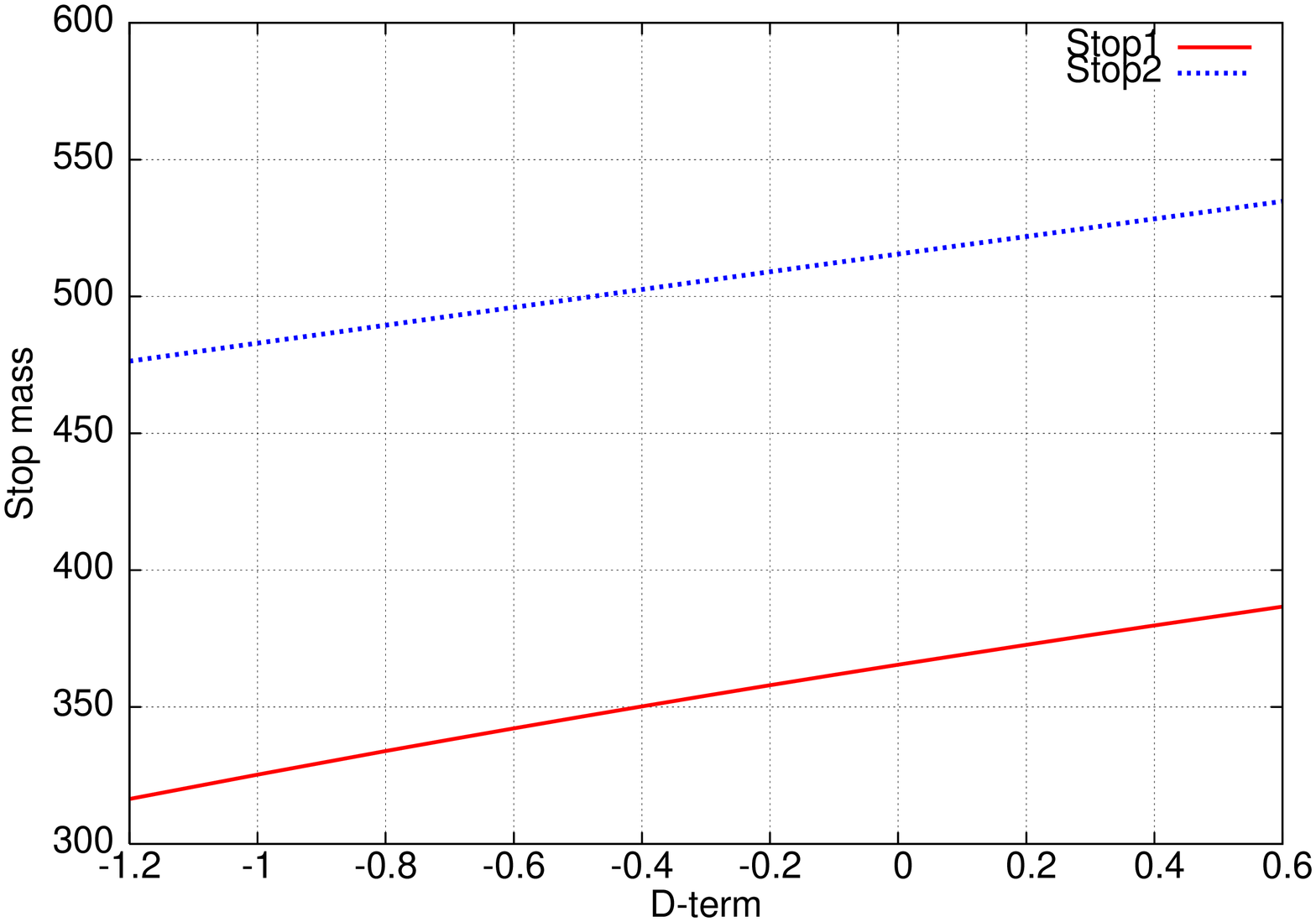,width=5.5cm,height=6.50cm,angle=-90}
\hskip 20pt \epsfig{file=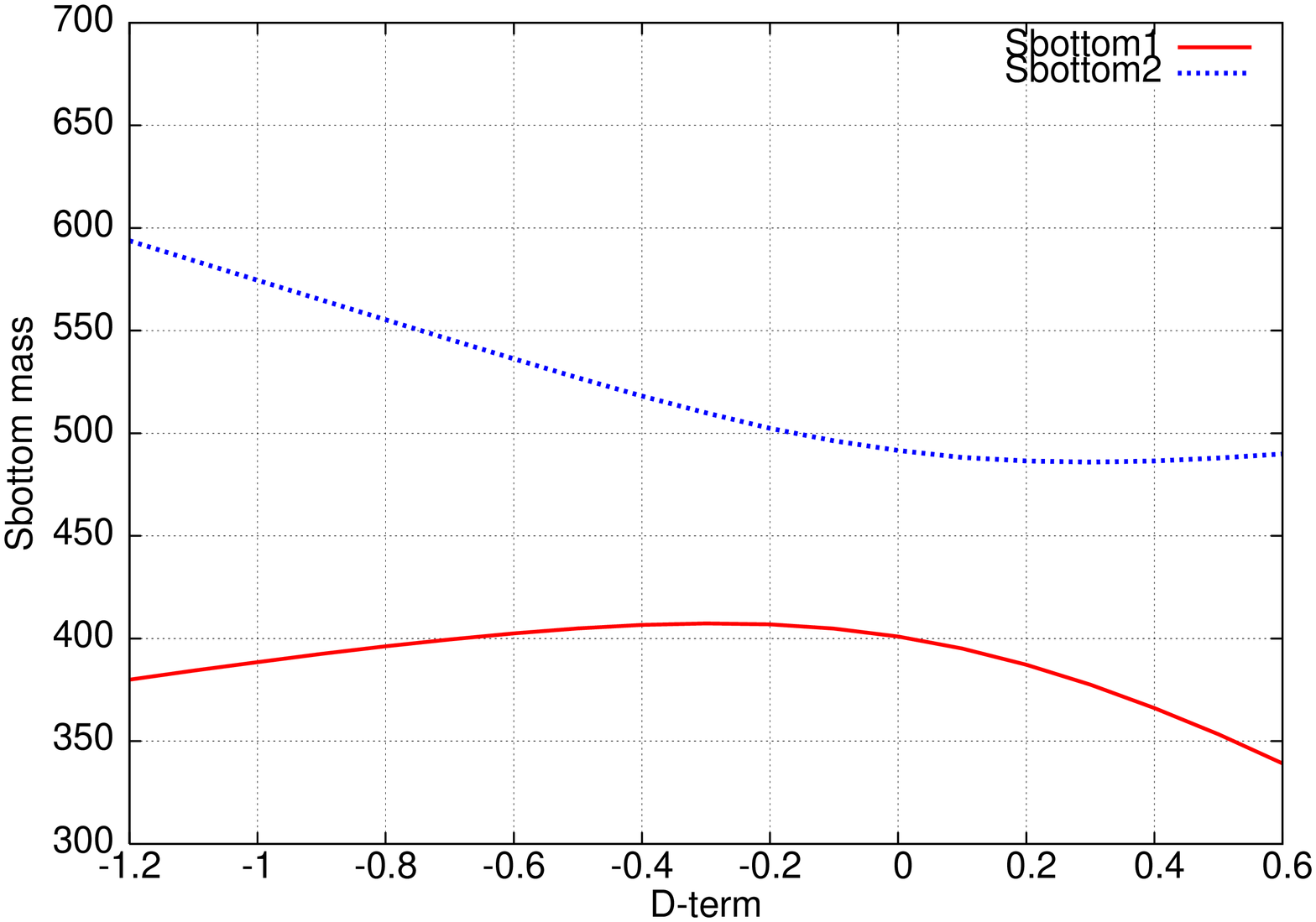,width=5.5cm,height=6.50cm,angle=-90}}
\caption{Variation of stop and sbottom mass with $D$-term:$\tan{\beta}=40$ } 
\end{center}

\end{figure}

\begin{figure}[t]
\begin{center}
\centerline{\epsfig{file=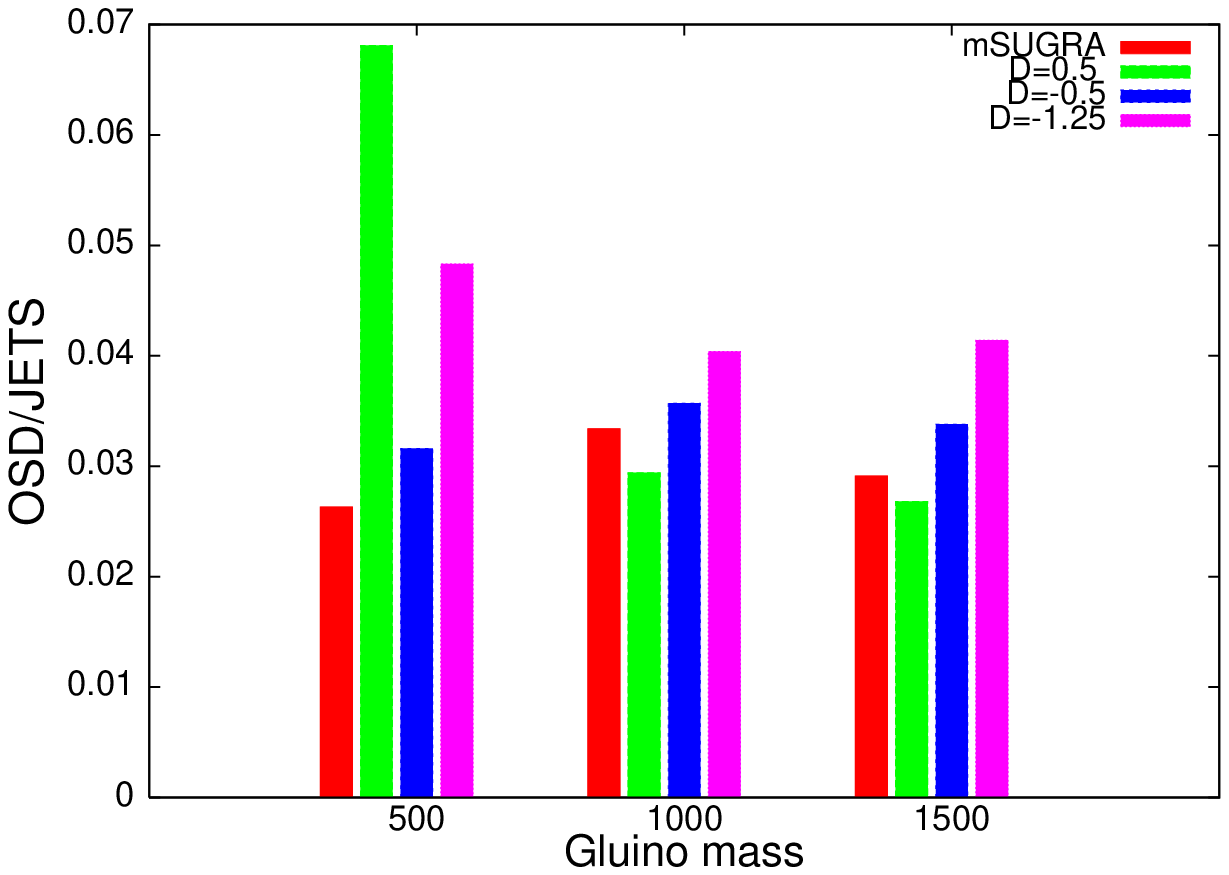,width=6.5 cm,height=5.50cm,angle=-0}
\hskip 20pt \epsfig{file=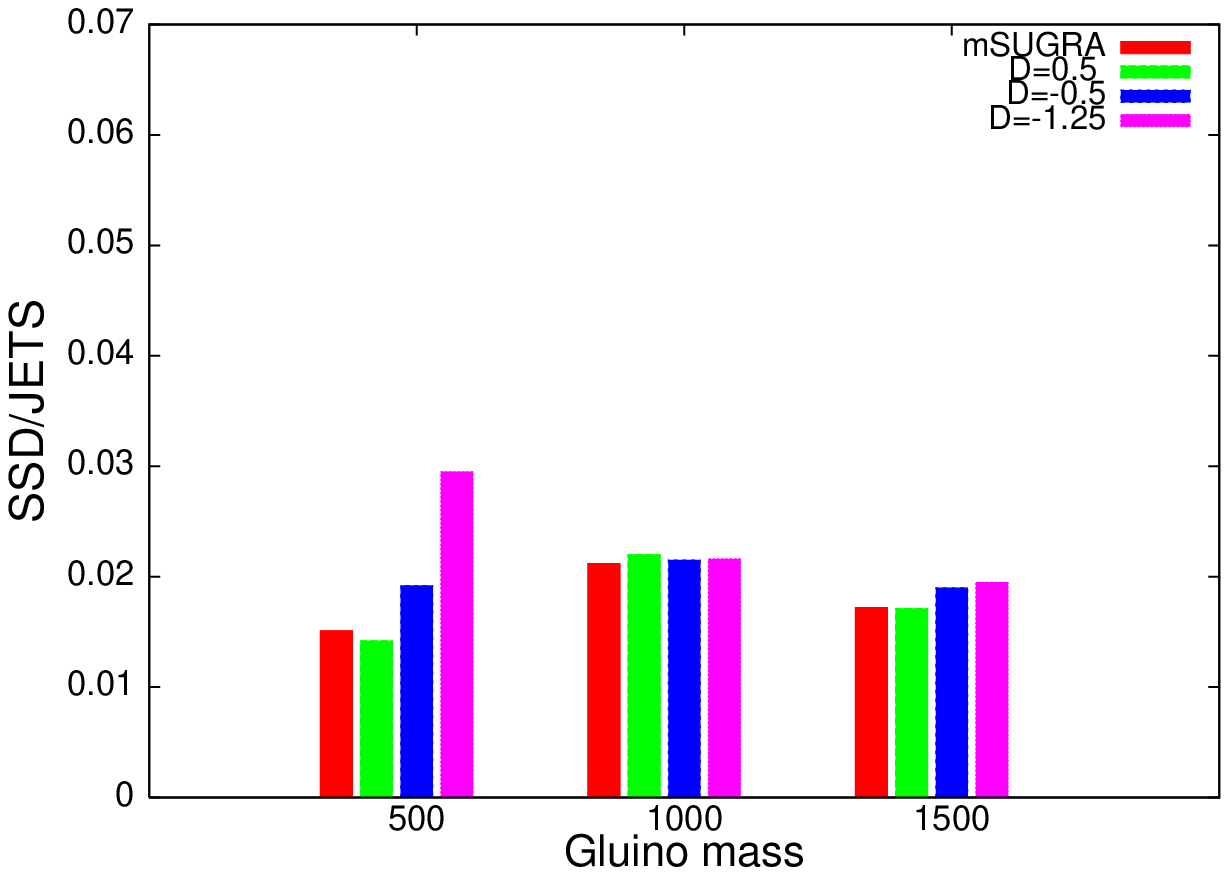,width=6.5cm,height=5.50cm,angle=-0}}
\vskip 10pt
\centerline{\epsfig{file=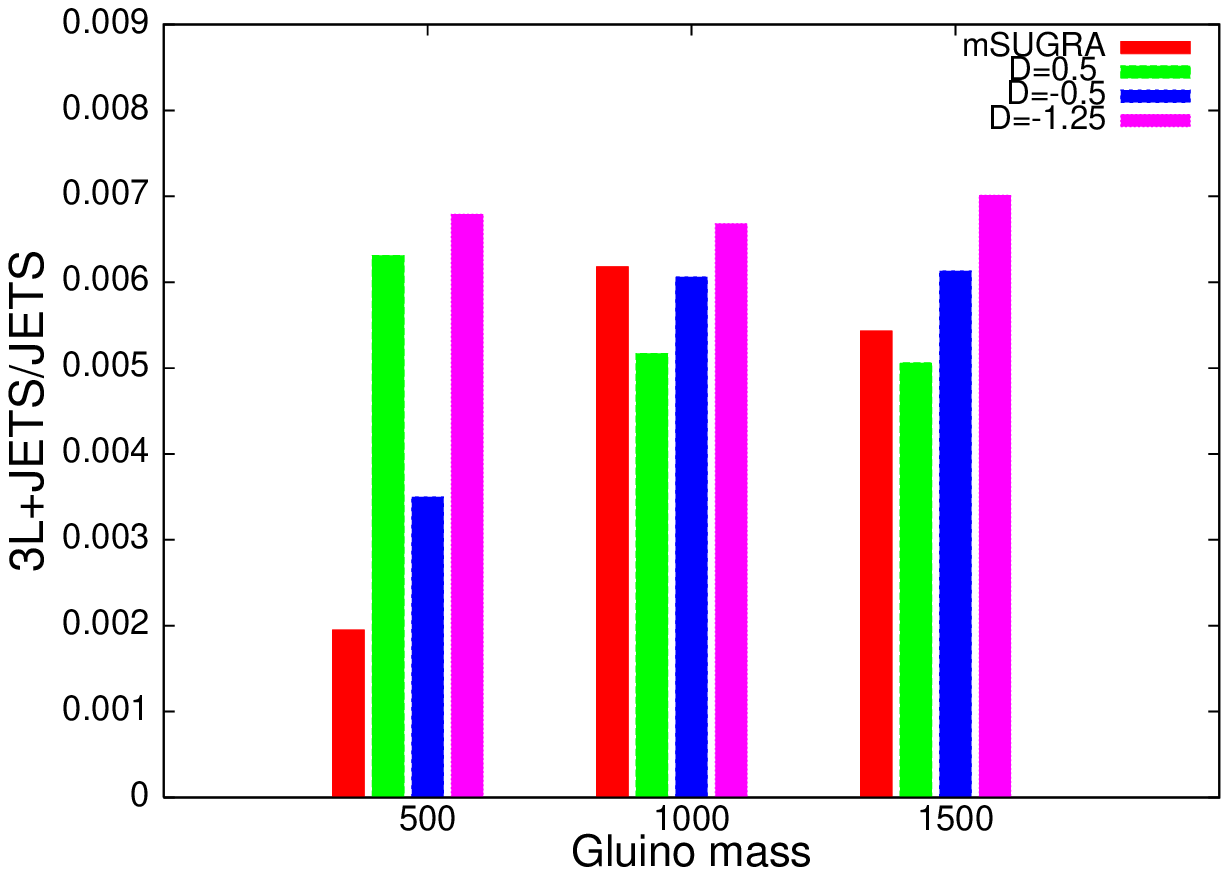,width=6.5 cm,height=5.50cm,angle=-0}
\hskip 20pt \epsfig{file=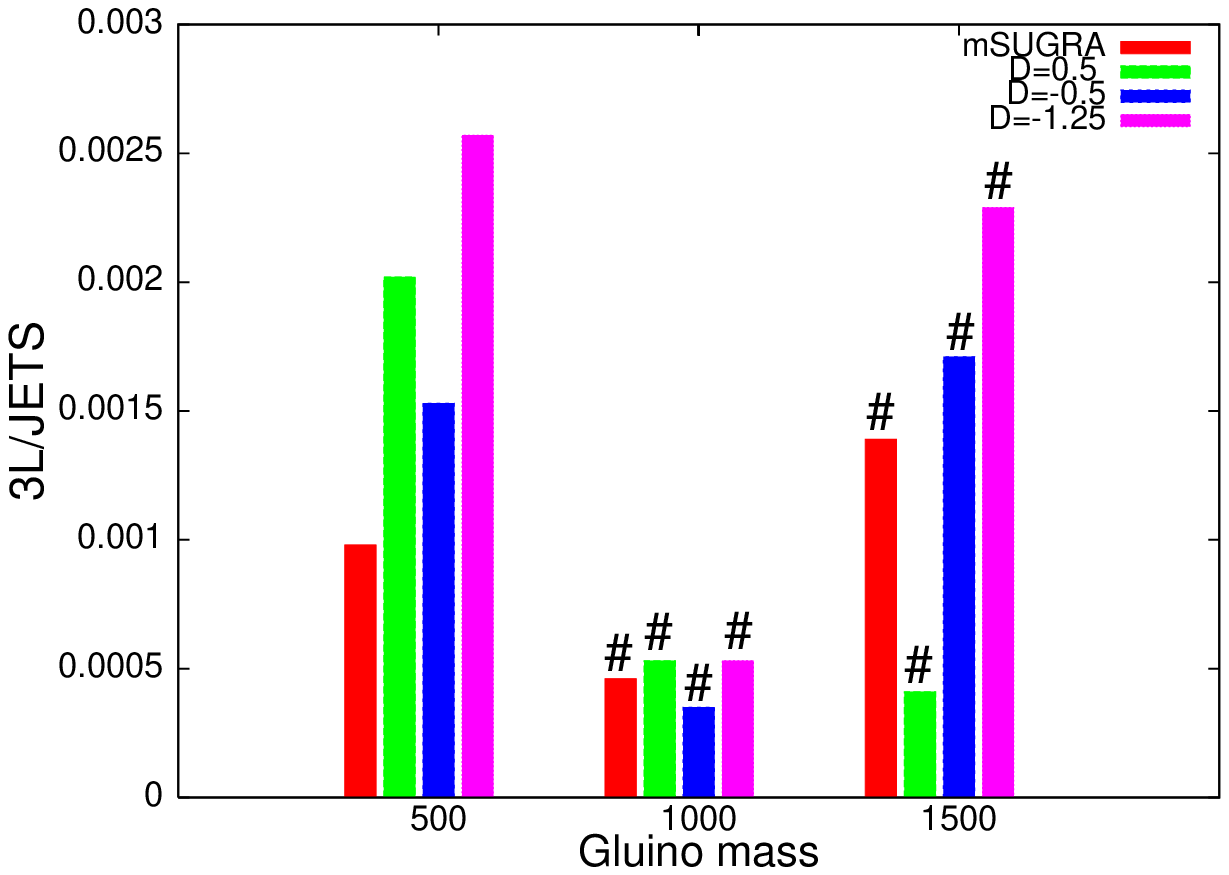,width=6.5cm,height=5.50cm,angle=-0}}
\caption{ Event ratios for $SO(10)~ D$-term Non-universality: $\tan{\beta}=5$} 
\end{center}

\end{figure}

\begin{figure}[t]
\begin{center}
\centerline{\epsfig{file=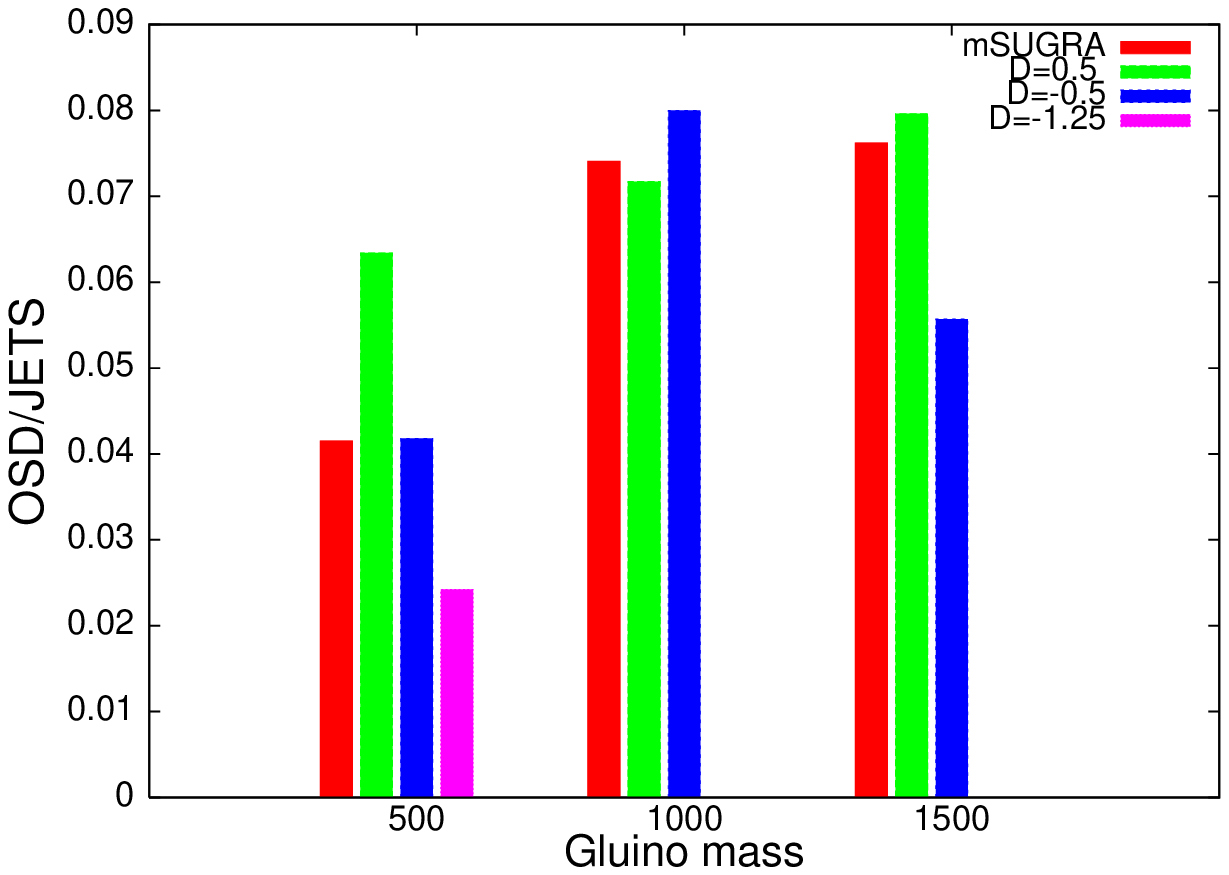,width=6.5 cm,height=5.50cm,angle=-0}
\hskip 20pt \epsfig{file=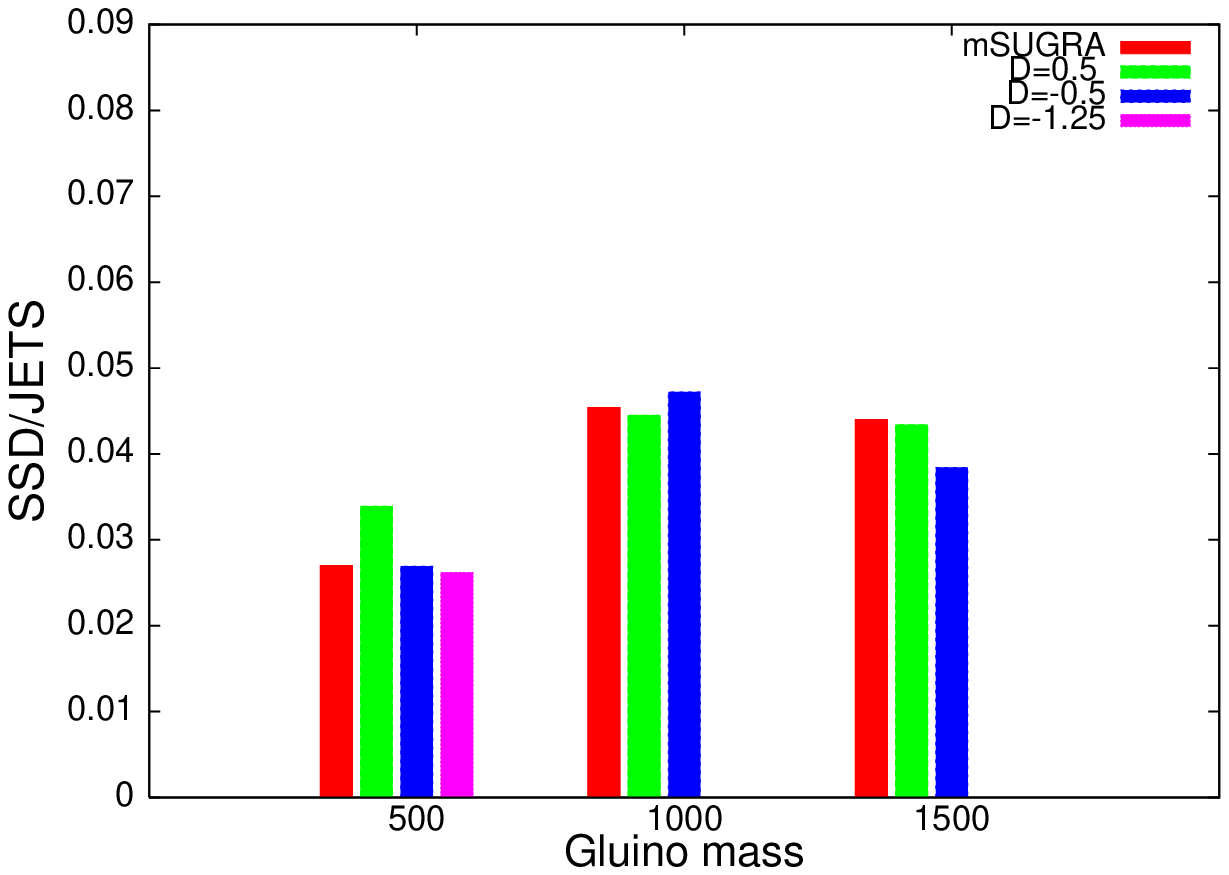,width=6.5cm,height=5.50cm,angle=-0}}
\vskip 10pt
\centerline{\epsfig{file=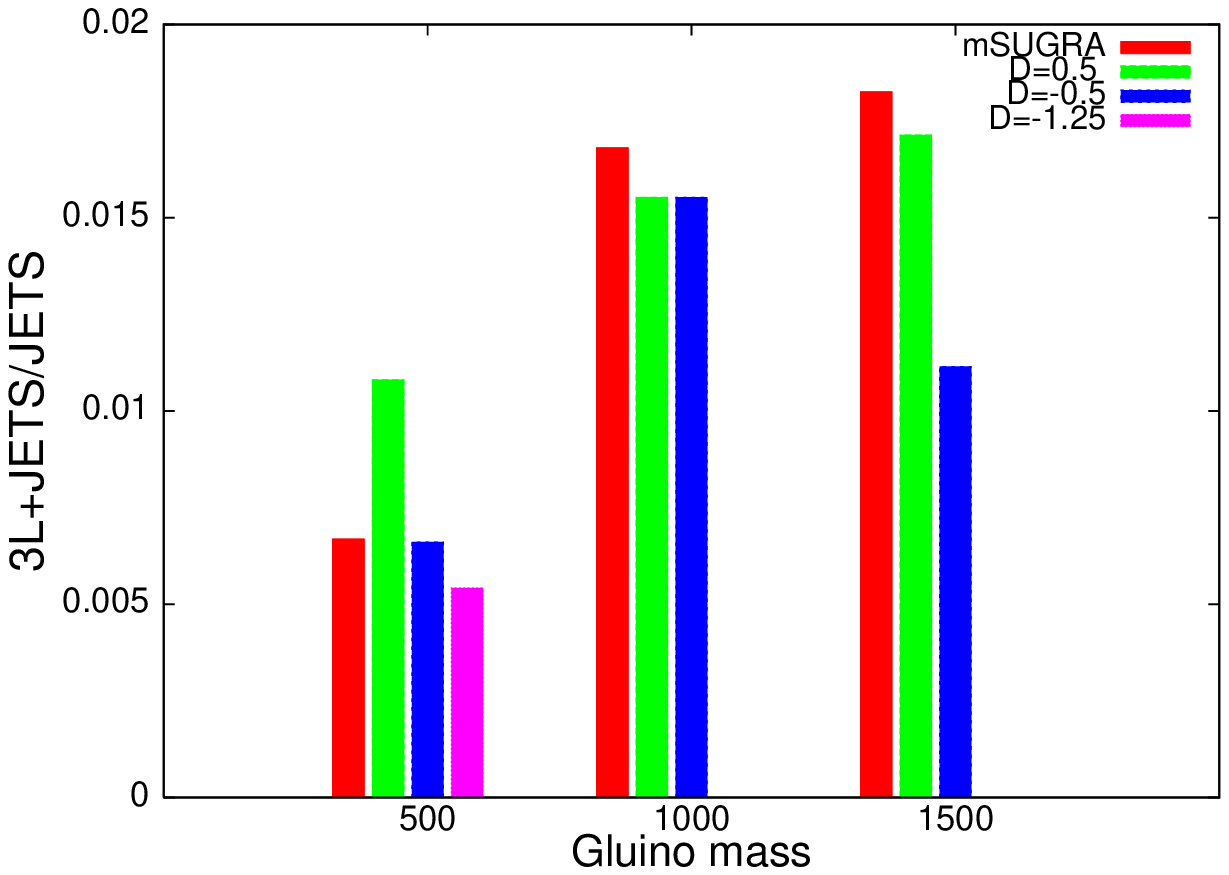,width=6.5 cm,height=5.50cm,angle=-0}
\hskip 20pt \epsfig{file=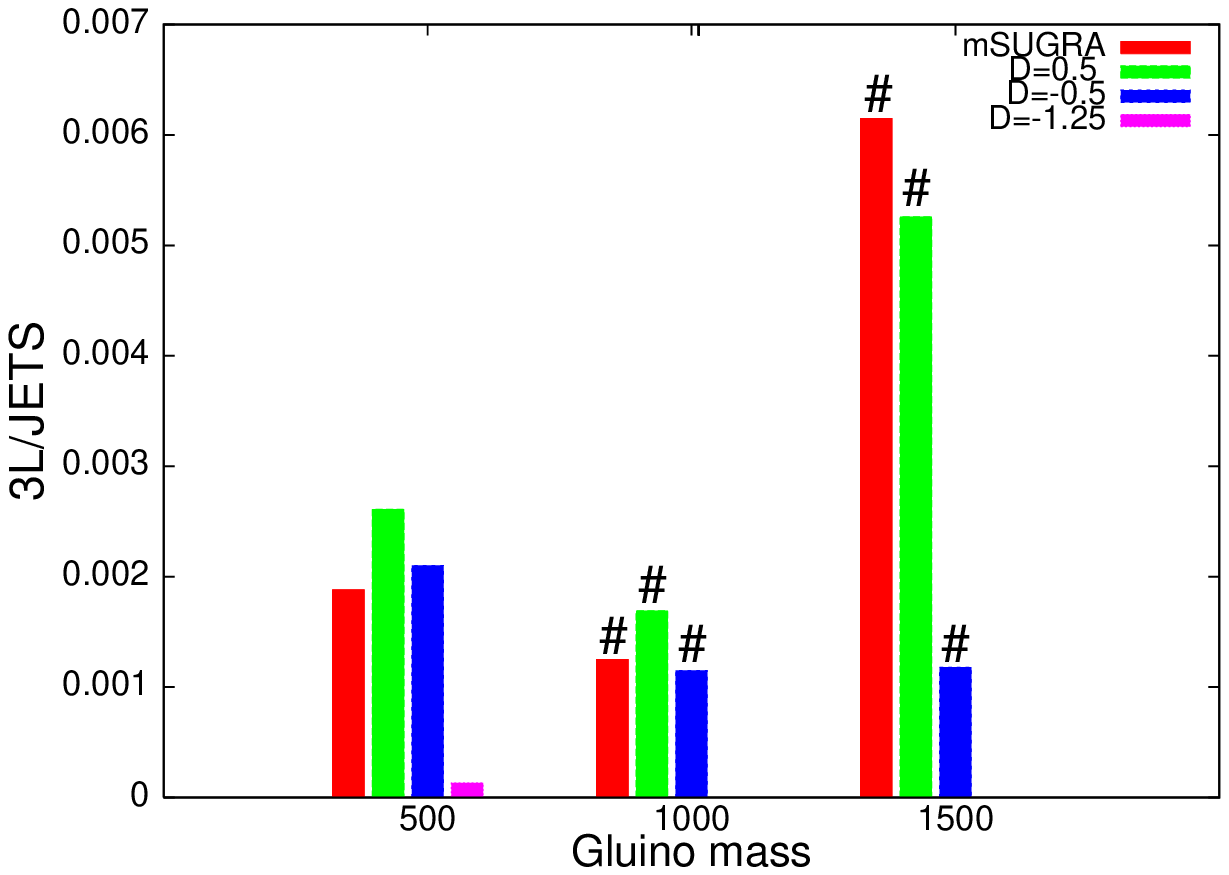,width=6.5cm,height=5.50cm,angle=-0}}
\caption{ Event ratios for $SO(10)~D$-term Non-universality: $\tan{\beta}=40$} 
\end{center}

\end{figure}

\subsection{Numerical results}

The low-energy masses of the right-chiral down-type squarks and
left-chiral charged leptons fall as $D$ is varied from the
minimum to the maximum allowed value in the permissible range. 
In particular, the masses of the physical states 
in the third family as a function
of $D$ are shown in figure 5 for both $\tan \beta$= 5 and 40, 
because they bring in more complex
behaviour due to mixing. It should be noted that the parabolic
$D$-dependence of the masses are flattened out considerably due to
running, since gauginos contribute to the low-energy scalar masses 
\cite{Rammond}.
The two stop mass eigenstates vary in the same way with $D$, since
both the $t_L$ and $t_R$ superfields belong to {\bf 10} of $SU(5)$,
while $b_R$, unlike $b_L$, belongs to ${\bf\bar{5}}$. The last mentioned
effect is responsible for different variation patterns of the
two sbottom mass eigenstates.

In any case, the nature of non-universality is different from
the two cases investigated earlier.

The same ratios as those studied previously are presented in
this context, in figures 6 and 7. The three values of $D$ mentioned 
above lead to the three non-universal bar graphs in each
case,  $D$ =0 being the corresponding
mSUGRA scenario. It may be noted that for $D$ = -1.25, one ends
up with a stau LSP for $m_{\tilde g}$ = 1 TeV, 1.5 TeV and $\tan\beta$ = 40.
The reason this does not happen for $m_{\tilde g}$ = 500 GeV 
is because the lowering of the lighter mass eigenstate is stalled by the
low value of $\mu$ in the first case.

The main features that emerge from the
ratio as well as the absolute rates are as follows:

\begin{itemize}

\item
For high gluino masses such as $m_{\tilde g}$= 1 TeV and 1.5 TeV, 
the distinction between various 
non-universalities for $D$= 0.5, -0.5 and -1.25 becomes 
difficult from the ratio plot.
This is because, for high value of $M_{1/2}$, the 
low energy squark-slepton masses are dominated by gaugino contributions, 
the effect of non-universal inputs to the scalar masses through 
$D$-terms being thus imperceptible. An exception to this occurs for
$m_{\tilde g}$= 1.5 TeV and $\tan \beta$ = 40, due to the same reason
as above, namely, the contribution to the off-diagonal
term in the sbottom mass matrix through the $\mu$-parameter
determined by such gaugino masses.

\item 
For $m_{\tilde g}$= 1 TeV and 1.5 TeV (particularly for
$\tan \beta$ =5), the only channels 
that partly distinguish among various values of 
the $SO(10)~ D$-term are $3\ell+jets$. 
This happens because whatever  mass hierarchy between squarks
and sleptons due to the $D$-terms is there is accentuated
with the largest detectable number of leptons in the final state.

\item
For $m_{\tilde g}$= 500 GeV, $OSD/jets$ is a good discriminator along with the 
trileptonic channels. In particular, the cases of $D$=0.5  and $D$=-1.25 are
 easily distinguishable from the ratios. The ratio $SSD/jets$, on the
other hand, is relatively flat, because these are initiated by
the production of gluinos, where the effects of scalars are more
often washed out.

\item The hadronically quiet trilepton events are largely washed out
by backgrounds, excepting for $m_{\tilde g}$= 500 GeV.

\item For $D$=-1.25, the leptonic final states give almost always the
largest  fraction of events for $\tan \beta$ = 5, while for 
$\tan \beta$ = 40 the fraction is the smallest.

\item
The absolute numbers in various channels are 
also very efficient discriminators in this type of 
non-universal scenarios particularly for low gluino mass 
(see table B5 and B6). 
\end{itemize}

\section{Summary and conclusions}

We have considered three representative scenarios where the scalar
mass spectrum in SUSY can deviate from the predictions of a universal
SUGRA model. These are situations with (a) high-scale non-universality
of squarks and sleptons, (b) a separate high-scale mass parameter for
the third family sfermions, and (c) the effect of $SO(10)~ D$-terms.
In each case, we have made a detailed scan of the parameter space,
in terms of the gluino and squark masses which set the scale of 
the hard scattering leading to superparticle production. While the 
value of the $\mu$ parameter (upto a sign) has been mostly fixed from 
radiative electroweak symmetry breaking, we have chosen two representative 
values of $\tan\beta$ for our analysis, namely 5 and 40. 

In essence, relatively low values of slepton 
masses in various schemes and in different regions of the parameter 
space buttress the leptonic final states.
With this in view, a multichannel analysis 
including various leptonic final states has been performed in each case, 
comparing the different degrees of non-universality with the mSUGRA 
case. The ratios of the like- and opposite-sign dilepton rates
as well as trileptons (with and without accompanying hard jets)
with respect to the inclusive jet signal. 

The case where the most conspicuous effects are seen in 
terms of the ratios is one where the the first two family squarks
have masses on the order of 10 TeV. In addition, the absolute number 
of events for this situation is rather low compared to the other cases,
which can serve as another distinguishing feature. 

For the first two family squarks still within 1 TeV or so, 
however, the distinction with the case of squark-slepton 
universality gets somewhat blurred. This is because the masses 
of the first two families of squarks and sleptons are often in the
same range, and thus the cascades leading to the leptonic
final states are similar in nature. A marginal, though not
spectacular, improvement is achieved by considering the 
absolute event rates. However, the ratios are more sensitive to
the mass ranges of the squarks and gluinos within a given 
pattern of non-universality, and as such they can provide 
useful clues to the level at which a departure from
universality has taken place.
The distinction is even more difficult for $SO(10)~ D$-terms,
except for $D$ = -1.25. For these  values of $D$, distinction
among various cases as well as with the universal case can
be problematic.

The effect of $\tan\beta$ can also have important bearing on the
various ratios an exception being in case of third family non-universality. 
Therefore, the independent extraction of  $\tan\beta$
from Higgs boson signals is going to be useful in establishing
scalar non-universality.

It is also seen that the trilepton events can be most useful
in making distinction among different situations. So are hadronically
quiet trleptons, so long as they are able to rise above backgrounds.
Next in the order is the importance of opposite-sign dileptons. Thus the 
investigation of leptonic final states with increasing
multiplicity, apart from the enhancing `clean' character 
of the events, is likely to enlighten us on the issue
of non-universality.

In addition to the different kinds of sfermion non-universality
discussed in the previous sections, one could also think of
the Higgs mass parameters evolving from a different common
high-scale value compared to that determining the squark and
slepton masses \cite{nonuH}. While this can affect Higgs phenomenology
considerably, our multichannel analysis gets appreciably
affected by such non-universality only when the charged Higgs
state can be made very light. In such case, too, the rates in
leptonic channels which are our main concern are altered if
the charged Higgs can be produced on-shell in the decay
of the stop or the sbottom , or of a chargino/neutralino.
Although the charged Higgs mass is lowered around or
below 200 GeV for some combinations of parameters
including a large $\tan \beta$, effects of the
above type are rare. 

It should be noted at the end that, unlike in the
case of gaugino non-universality \cite{Subho}, the schemes of
parametrising scalar non-universality are more
non-uniform. Therefore, different schemes often lead
to overlapping portions in the spectrum, where
signals may turn out to be of similar nature.
The most significant departure from universality in terms 
of overall event rates can occur through the variation
of masses of the first two family squarks, whereas
the lepton-to-jet event ratios are influenced
more substantially when the first two family sleptons
have masses that are different from what is predicted
in mSUGRA. These generic features of the scalar
spectrum, rather than different theoretical schemes,
are likely to be exposed more easily at the LHC.

\noindent
\acknowledgments We thank Howard Baer, Priyotosh Bandyopadhyay 
and Sudhir Kumar Gupta for
useful suggestions. Our work was partially supported by funding available 
from the Department of Atomic Energy, Government of India, 
for the Regional Centre for Accelerator-based
Particle Physics, Harish-Chandra Research Institute. 
Computational work for this study was
partially carried out at the cluster computing facility in the
Harish-Chandra Research Institute ({\tt http:/$\!$/cluster.mri.ernet.in}).

\newpage
\noindent
{\large {\bf APPENDIX A}}\\

\noindent
In this appendix we list the relevant masses in the spectrum. Specifically,
we provide the high scale scalar inputs (which is specific to the kind of 
non-universal model) to generate the low energy scalar 
mass parameters. We provide the low lying 
chargino-neutralino masses as well. The tables are organised as follows:
 squark-Slepton non-universal case in A1 and A2, 
third generation scalar non-universality and in A3 and A4, and 
non-universality arising due to $SO(10)~ D$-term in A5 and A6.

We would like to mention that for low energy $m_{\tilde g}$= 500 GeV, 
1000 GeV, or 1500 GeV, high scale universal input for the gaugino masses 
$m_{ 1/2}$ are 166.9 GeV, 333.65 GeV and 500.5 GeV for 1-loop RGE and 
this is obviously independent of what kind of scalar non-universal model 
we are looking at.

NA indicates that the spectrum generated is inconsistent 
due to the reasons mentioned in the text accordingly.
\begin{center}
\begin{tabular}{|c|c|c|c|c|c|c|c|c|c|c|}
\multicolumn{11}{c}{Table A1 : Mass Spectrum (GeV) for 
squark-slepton non-universality }\\
\multicolumn{11}{c}{$\tan \beta$= 5} \\
\multicolumn{11}{c}{(Figure 1 )}\\
\hline
 ($m_{\tilde g}$,$m_{\tilde {q}^{1,2}}$) & $m_{\tilde {l}^{1,2}}$ 
& $m_{0\tilde q}$ & 
$m_{0\tilde l}$ &$m_{\tilde \chi_{2}^{\pm}}$ & $m_{\tilde \chi_{1}^{\pm}}$ & 
$m_{\tilde \chi_{2}^{0}}$ & $m_{\tilde \chi_{1}^{0}}$ & 
$m_{\tilde t_{1}}$ & $m_{\tilde b_{1}}$ & $m_{\tilde \tau_{1} }$\\
\hline 
(500,500) &{\bf 225* } & 200 & 200 & 344 & 117 & 118 & 60 & 336 & 450 & 212 \\
\hline
(500,500) &{\bf 250} & 200 & 220 & 337 & 116 & 117 & 60 & 333 & 449 & 231 \\
\hline
(500,500) &{\bf 500} & NA & NA & NA & NA & NA & NA & NA & NA & NA \\
\hline
(500,500) &{\bf 750} & NA & NA & NA & NA & NA & NA & NA & NA & NA \\
\hline
(500,1000) &{\bf 906*} & 900 & 900 & 546 & 126 & 127 & 62 & 558 & 818 & 900\\
\hline
(500,1000) &{\bf 250} & 900 & 230 & 867 & 130 & 130 & 63 & 705 & 876 & 234 \\
\hline
(500,1000) &{\bf 500} & 900 & 490 & 799 & 130 & 130 & 63 & 672 & 862 & 491  \\
\hline
(500,1000) &{\bf 750} & 900 & 740 & 674 & 128 & 128 & 62 & 613 & 838 & 740  \\
\hline
(1000,1000) &{\bf 450*} & 400 & 400 & 668 & 259 & 259 & 126 & 709 & 896 & 421 \\
\hline
(1000,1000) &{\bf 250} & 400 & 0 & 736 & 261 & 261 & 126 & 734 & 907 & 133 \\
\hline
(1000,1000) &{\bf 500} & 400 & 431 & 657 & 259 & 259 & 126 & 705 & 894 & 450 \\
\hline
(1000,1000) &{\bf 750} & 400 & 705 & 499 & 252 & 252 & 125 & 652 & 871 & 716 \\
\hline
\multicolumn{11}{c}{{\bf * marked cases correspond to mSUGRA}}\\
\multicolumn{11}{c}{ ($m_{0\tilde q}$ and  $m_{0\tilde l}$ 
are high scale non-universal inputs of squark and slepton mass)}
\end {tabular}
\end {center}

\begin{center}
\begin{tabular}{|c|c|c|c|c|c|c|c|c|c|c|}
\multicolumn{11}{c}{Table A2 : Mass Spectrum (GeV)for 
squark-slepton non-universality}\\
\multicolumn{11}{c}{$\tan \beta$= 40} \\
\multicolumn{11}{c}{(Figure 2 )}\\
\hline
 ($m_{\tilde g}$,$m_{\tilde {q}^{1,2}}$) & $m_{\tilde {l}^{1,2}}$ 
& $m_{0\tilde q}$ & 
$m_{0\tilde l}$ &$m_{\tilde \chi_{2}^{\pm}}$ & $m_{\tilde \chi_{1}^{\pm}}$ & 
$m_{\tilde \chi_{2}^{0}}$ & $m_{\tilde \chi_{1}^{0}}$ & 
$m_{\tilde t_{1}}$ & $m_{\tilde b_{1}}$ & $m_{\tilde \tau_{1} }$\\
\hline 
 (500,500) &{\bf 225*} & 200 & 200 & 320 & 122 & 122 & 62 & 344 & 371 & 134\\
\hline
 (500,500) &{\bf 250} & 200 & 220 & 312 & 121 & 122 & 62 & 341 & 371 & 156 \\
\hline
 (500,500) &{\bf 500} & NA & NA & NA & NA & NA & NA & NA & NA & NA \\
\hline
(500,500) &{\bf 750} & NA & NA & NA & NA & NA & NA & NA & NA & NA\\
\hline
(500,1000) &{\bf 906* } & 900 & 900 & 457 & 129 & 129 & 63 & 578 & 684 & 731\\
\hline
(500,1000) &{\bf 250} & 900 & 230 & 827 & 132 & 132 & 63 & 719 & 750 & 921\\
\hline
(500,1000) &{\bf 500} & 900 & 490 & 752 & 132 & 132 & 63 & 687 & 734 & 381 \\
\hline
(500,1000) &{\bf 750} & 900 & 740 & 611 & 131 & 131 & 63 & 631 & 707 & 599  \\
\hline
(1000,1000) &{\bf 450*} & 400 & 400 & 620 & 262 & 262 & 127 & 718 & 788 & 317\\
\hline
(1000,1000) &{\bf 250} & NA & NA & NA & NA & NA & NA & NA & NA & NA \\
\hline
(1000,1000) &{\bf 500} & 400 & 431 & 607 & 262 & 262 & 127 & 714 & 787 & 344 \\
\hline
(1000,1000) &{\bf 750} & 400 & 705 & 423 & 251 & 251 & 126 & 661 & 769 & 572 \\
\hline
\multicolumn{11}{c}{{\bf * marked cases correspond to mSUGRA}}\\
\multicolumn{11}{c}{ ($m_{0\tilde q}$ and  $m_{0\tilde l}$ 
are high scale non-universal inputs of squark and slepton mass)}
\end{tabular}
\end{center}

\begin{center}
\begin{tabular}{|c|c|c|c|c|c|c|c|c|c|c|}
\multicolumn{11}{c}{Table A3 : Mass Spectrum(GeV) for 
Third family scalar non-universality }\\
\multicolumn{11}{c}{$\tan \beta$= 5} \\
\multicolumn{11}{c}{(Figure 3 )}\\
\hline
 ($m_{\tilde g}$,$m_{\tilde t_{1}}$) & $m_{\tilde q^{1,2}}$ & $m^3_{0}$ & 
$m^{(1,2)}_{0}$ &$m_{\tilde \chi_{2}^{\pm}}$ & $m_{\tilde \chi_{1}^{\pm}}$ & 
$m_{\tilde \chi_{2}^{0}}$ & $m_{\tilde \chi_{1}^{0}}$ & 
$m_{\tilde {l}^{1,2}}$ & $m_{\tilde \tau_{1}}$ & $m_{\tilde b_{1}}$ \\
\hline 
 (500,500) &{\bf 876* } & 750 & 750 & 490 & 125 & 125 & 62 & 758 & 751 & 720\\
\hline
 (500,500) &{\bf 1000} & 750 & 900 & 490 & 125 & 125 & 62 & 906 & 751 & 720\\
\hline
 (500,500) &{\bf 10000} & 750 & 9990 & 490 & 125 & 125 & 62 & 9990 &751 & 720\\
\hline
(500,1000) &{\bf 2050*} & 2000 & 2000 & 1024 & 131 &131 &63 &2000 &1995 &1611\\
\hline
(500,1000) &{\bf 1000} & 2000 & 900 & 1024 & 131 & 131 & 63 &906 & 1995 &1611\\
\hline
(500,1000) &{\bf 10000} & 2000 & 9990 & 1024 & 131 &131 &63 &9990 &1995 &1611\\
\hline
(1000,1000) &{\bf 1765*} & 1510 & 1510 &973 &263 &263 &126 & 1525 &1512 &1444\\
\hline
(1000,1000) &{\bf 1000} & 1510 & 400 & 973 & 263 &263 &126 &450 &1512 & 1444\\
\hline
(1000,1000) &{\bf 10000} & 1510 & 9990 & 973 &263 &263 &126 &9990 &1512 &1444\\
\hline
\multicolumn{11}{c}{{\bf * marked cases correspond to mSUGRA}}\\
\multicolumn{11}{c}{ ($m^3_{0}$ and $m^{(1,2)}_{0}$ 
are high scale inputs of 3rd and 1,2 family non-universal scalar mass)}\\
\end{tabular}
\end{center}

\begin{center}
\begin{tabular}{|c|c|c|c|c|c|c|c|c|c|c|}
\multicolumn{11}{c}{Table A4 : Mass Spectrum(GeV) for 
Third family scalar non-universality }\\
\multicolumn{11}{c}{$\tan \beta$= 40} \\
\multicolumn{11}{c}{(Figure 4 )}\\
\hline
 ($m_{\tilde g}$,$m_{\tilde t_{1}}$) & $m_{\tilde q^{1,2}}$ & $m^3_{0}$ & 
$m^{(1,2)}_{0}$ &$m_{\tilde \chi_{2}^{\pm}}$ & $m_{\tilde \chi_{1}^{\pm}}$ & 
$m_{\tilde \chi_{2}^{0}}$ & $m_{\tilde \chi_{1}^{0}}$ & 
$m_{\tilde {l}^{1,2}}$ & $m_{\tilde \tau_{1}}$ & $m_{\tilde b_{1}}$ \\
\hline 
 (500,500) &{\bf 876*} & 750 & 750 & 418 & 128 & 128 & 63 & 758 & 608 & 604  \\
\hline
 (500,500) &{\bf 1000} & 750 & 900 & 418 & 128 & 128 & 63 & 906 & 608 & 604 \\
\hline
 (500,500) &{\bf 10000} & 750 & 9990 & 418 & 128 & 128 & 63 & 9990 &608 &604 \\
\hline
(500,1000) &{\bf 2050*} & 2000 &2000 & 811 & 132 &132 &63 &2000 &1626 &1331 \\
\hline
(500,1000) &{\bf 1000} & 2000 & 900 & 811 & 132 & 132 & 63 &906 &1626 &1331  \\
\hline
(500,1000) &{\bf 10000} & 2000 & 9990 & 811 & 132 &132 &63 &9990 &1626 &1331 \\
\hline
(1000,1000) &{\bf 1765*} & 1510 & 1510 & 827 &265 &265 &127 &1525 &1230 &1236\\
\hline
(1000,1000) &{\bf 1000} & 1510 & 400 & 827 & 265 &265 & 127 &450 &1230 &1236 \\
\hline
(1000,1000) &{\bf 10000} & 1510 & 9990 & 927 &265 &265 &127 &9990 &1230 &1236\\
\hline
\multicolumn{11}{c}{{\bf * marked cases correspond to mSUGRA}}\\
\multicolumn{11}{c}{ ($m^3_{0}$ and $m^{(1,2)}_{0}$ 
are high scale inputs of 3rd and 1,2 family non-universal scalar mass)}
\end{tabular}
\end{center}

\begin{center}
\begin{tabular}{|c|c|c|c|c|c|c|c|c|c|c|c|}
\multicolumn{12}{c}{Table A5 : Mass Spectrum(GeV) for 
$SO(10)~ D$-term scalar Non-universality }\\
\multicolumn{12}{c}{High scale scalar mass input $m_0$=300 GeV} \\
\multicolumn{12}{c}{$\tan \beta$= 5} \\
\multicolumn{12}{c}{(Figure 6 )}\\
\hline
 $m_{\tilde g}$ & D-term & $m_{\tilde \chi_{2}^{\pm}}$ & 
$m_{\tilde \chi_{1}^{\pm}}$ & $m_{\tilde \chi_{2}^{0}}$ & 
$m_{\tilde \chi_{1}^{0}}$ & $m_{\tilde e_{L}}$ & $m_{\tilde u_{L}}$ & 
$m_{\tilde d_{R}}$ & $m_{\tilde t_{1}}$ & $m_{\tilde b_{1}}$ 
& $m_{\tilde \tau_{1} }$\\
\hline 
 500 &{\bf 0.0*} & 361 & 118 & 119 & 60 &328 &548 &537 &358 &483 &308  \\
\hline
 500 &{\bf 0.5} & 386 & 120 & 121 &60 &201 &568 &470 &374 &467 &200\\
\hline
 500 &{\bf -0.5} & 334 & 116 & 117 & 59 &419 &527 &597 &342 &466 &270 \\
\hline
 500 &{\bf -1.25} & 291  & 110 & 112 &58 &526 &494 &676 &316 &438 &198 \\
\hline
 1000 &{\bf 0.0*} & 656 & 259 & 259 & 126 &394 &968 &939 &695 &872 &328 \\
\hline
 1000 &{\bf 0.5} & 670 & 259 & 259 & 126 &297 &980 &903 &703 &880 &287  \\
\hline
 1000 &{\bf -0.5} & 641 & 259 & 259 & 126 &472 &956 &975 &686 &863 &292  \\
\hline
 1000 &{\bf -1.25} & 619 & 258 & 258 & 126 &569 &939 &1025 &673 &849 &227  \\
\hline
 1500 &{\bf 0.0*} & 965 & 396 & 396 & 190 &485 &1414 &1368 &1052 &1281 &359 \\
\hline
 1500 &{\bf 0.5} & 975 & 396 & 396 & 190 &409 &1422 &1344 &1057 &1287 &384 \\
\hline
 1500 &{\bf -0.5} & 955 & 396 & 396 & 190 &550 &1406 &1393 &1047 &1275 &326  \\
\hline
 1500 &{\bf -1.25} & 940 & 396 & 396 & 190 &635 &1394 &1429 &1039 &1265 &270 \\
\hline
\multicolumn{12}{c}{{\bf * marked cases correspond to mSUGRA}}
\end {tabular}
\end {center}

\begin{center}
\begin{tabular}{|c|c|c|c|c|c|c|c|c|c|c|c|}
\multicolumn{12}{c}{Table A6 : Mass Spectrum(GeV) for 
$SO(10)~ D$-term scalar Non-universality }\\
\multicolumn{12}{c}{High scale scalar mass input $m_0$=300 GeV} \\
\multicolumn{12}{c}{$\tan \beta$= 40} \\
\multicolumn{12}{c}{(Figure 6 )}\\
\hline
 $m_{\tilde g}$ & D-term & $m_{\tilde \chi_{2}^{\pm}}$ & 
$m_{\tilde \chi_{1}^{\pm}}$ & $m_{\tilde \chi_{2}^{0}}$ & 
$m_{\tilde \chi_{1}^{0}}$ & $m_{\tilde e_{L}}$ & $m_{\tilde u_{L}}$ & 
$m_{\tilde d_{R}}$ & $m_{\tilde t_{1}}$ & $m_{\tilde b_{1}}$ 
& $m_{\tilde \tau_{1} }$\\
\hline 
 500 &{\bf 0.0*} & 330 & 123 & 123 & 62 &329 &547 &537 &365 &401 &229  \\
\hline
 500 &{\bf 0.5} & 358 & 125 & 125 & 62 &202 &568 &470 &383 &353 &139  \\
\hline
 500 &{\bf -0.5} & 301 & 120 & 121 & 62 &419 &526 &597 &346 &405 &184 \\
\hline
 500 &{\bf -1.25} & 252 & 112 & 113 & 60 &526 &493 &676 &314 &378 &305  \\
\hline
 1000 &{\bf 0.0*} & 612 & 262 & 262 & 127 &395 &968 &940 &704 &767 &229\\
\hline
 1000 &{\bf 0.5} & 628 & 262 & 262 & 127 &297 &980 &903 &713 &746 &204  \\
\hline
 1000 &{\bf -0.5} & 596 & 261 & 261 & 127 &472 &956 &975 &694 &775 &189  \\
\hline
 1000 &{\bf -1.25} & NA & NA & NA & NA &NA &NA &NA &NA &NA &NA  \\
\hline
 1500 &{\bf 0.0*} & 905 & 398 & 398 & 191 &485 &1414 &1369 &1065 &1151 &250 \\
\hline
 1500 &{\bf 0.5} & 916 & 398 & 398 & 191 &409 &1422 &1344 &1071 &1138 &259 \\
\hline
 1500 &{\bf -0.5} & 894 & 398 & 398 & 191 &550 &1406 &1393 &1059 &1159 &211  \\
\hline
 1500 &{\bf -1.25} & NA & NA & NA & NA &NA &NA &NA &NA &NA &NA  \\
\hline
\multicolumn{12}{c}{{\bf * marked cases correspond to mSUGRA}}
\end {tabular}
\end {center}

\newpage
\noindent
{\large {\bf APPENDIX B}}

Here we provide cross sections for all the channels in the three
 non-universal scenarios studied a) Squark-Slepton Non-universal case, 
b) 3rd generation scalar non-universality and c) Non-universality arising 
due to $SO(10)~ D$-term respectively in three tables a) B1, B2 b) B3, B4
c) B5, B6. The SM background cross section is tabulated in B7.

The cross-sections are named as follows:
 $\sigma_{OSD}$ for OSD, $\sigma_{SSD}$ for SSD, $\sigma_{3\ell+jets}$ 
for $(3\ell+jets)$, $\sigma_{(3\ell)}$ for $(3\ell)$ and 
$\sigma_{jets}$ for $jets$.

The cross-sections in bold font indicate that it is submergerd 
in the background as defined in text.

NA indicates that the spectrum is inconsistent as discussed early.
%
\begin{center}
\begin{tabular}{|c|c|c|c|c|c|c|}
\multicolumn{7}{c}{Table B1 : Cross-sections (pb) for 
squark-slepton non-universality}\\
\multicolumn{7}{c}{$\tan \beta$= 5} \\
\multicolumn{7}{c}{(Figure 1 )}\\
\hline
 ($m_{\tilde g}$,$m_{\tilde q^{1,2}}$) & $m_{\tilde l^{1,2}}$ 
& $\sigma_{OSD}$ & $\sigma_{SSD}$ &$\sigma_{(3\ell+jets)}$ 
& $\sigma_{(3\ell)}$ & $\sigma_{jets}$ \\
\hline 
 (500,500) &{\bf mSUGRA} & 0.4972 & 0.2100 & 0.0437 & 0.00111 & 9.3302 \\
\hline
 (500,500) &{\bf 250} & 0.4144 & 0.2316 & 0.0367 & 0.01836 & 10.351 \\
\hline
 (500,500) &{\bf 500} & NA & NA & NA & NA & NA  \\
\hline
(500,500) &{\bf 750} & NA & NA & NA & NA & NA  \\
\hline
(500,1000) &{\bf mSUGRA} & 0.1782 & 0.0948 & 0.0266 & 0.00224 & 7.1574 \\
\hline
(500,1000) &{\bf 250} & 0.5218 & 0.1526 & 0.0931 & 0.01357 & 7.3764 \\
\hline
(500,1000) &{\bf 500} & 0.2989 & 0.1019 & 0.0440 & 0.00380 & 7.3032  \\
\hline
(500,1000) &{\bf 750} & 0.1593 & 0.0955 & 0.0231 & 0.00220 & 7.2698  \\
\hline
(1000,1000) &{\bf mSUGRA} & 0.0277 & 0.0185 & 0.0060 & \bf{0.00034} & 0.7277 \\
\hline
(1000,1000) &{\bf 250} & 0.0261 & 0.0186 & 0.0049 & \bf{0.00024} & 0.3838 \\
\hline
(1000,1000) &{\bf 500} & 0.0289 & 0.0193 & 0.0060 & \bf{0.00032} & 0.7285  \\
\hline
(1000,1000) &{\bf 750} & 0.0333 & 0.0231 & 0.0082 & \bf{0.00031} & 0.7851  \\
\hline
\end {tabular}
\end{center}
%
\begin{center}
\begin{tabular}{|c|c|c|c|c|c|c|}
\multicolumn{7}{c}{Table B2 : Cross-sections (pb) for 
squark-slepton non-universality}\\
\multicolumn{7}{c}{$\tan \beta$= 40} \\
\multicolumn{7}{c}{(Figure 2 )}\\
\hline
 ($m_{\tilde g}$,$m_{\tilde q^{1,2}}$) & $m_{\tilde l^{1,2}}$ 
& $\sigma_{OSD}$ & $\sigma_{SSD}$ &$\sigma_{(3\ell+jets)}$ 
& $\sigma_{(3\ell)}$ & $\sigma_{jets}$ \\
\hline 
 (500,500) &{\bf mSUGRA} & 0.6267 & 0.3466 & 0.0665 & 0.02215 & 14.0742 \\
\hline
 (500,500) &{\bf 250} & 0.5079 & 0.2971 & 0.0713 & 0.01585 & 14.4145 \\
\hline
 (500,500) &{\bf 500} & NA & NA & NA & NA & NA  \\
\hline
(500,500) &{\bf 750} & NA & NA & NA & NA & NA  \\
\hline
(500,1000) &{\bf mSUGRA} & 0.2388 & 0.1317 & 0.0441 & 0.00657 & 6.8736 \\
\hline
(500,1000) &{\bf 250} & 0.2730 & 0.1886 & 0.0422 & 0.00561 & 7.1379  \\
\hline
(500,1000) &{\bf 500} & 0.2798 & 0.1248 & 0.0556 & 0.00532 & 7.0394  \\
\hline
(500,1000) &{\bf 750} & 0.2037 & 0.1246 & 0.0319 & 0.00509 & 6.9650  \\
\hline
(1000,1000) &{\bf mSUGRA} & 0.0314 & 0.0203 & 0.0066 & \bf{0.00034} & 0.7839 \\
\hline
(1000,1000) &{\bf 250} & NA & NA & NA & NA & NA \\
\hline
(1000,1000) &{\bf 500} & 0.03323 & 0.0205 & 0.0066 & \bf{0.00036} & 0.7900  \\
\hline
(1000,1000) &{\bf 750} & 0.0393 & 0.0209 & 0.0093 & \bf{0.00068} & 0.8101  \\
\hline
\end {tabular}
\end {center}
%
%
\begin{center}
\begin{tabular}{|c|c|c|c|c|c|c|}
\multicolumn{7}{c}{Table B3 : Cross-sections (pb) for 
Third family scalar non-universality}\\
\multicolumn{7}{c}{$\tan \beta$= 5} \\
\multicolumn{7}{c}{(Figure 3 )}\\
\hline
 ($m_{\tilde g}$,$m_{\tilde q^{3}}$) & $m_{\tilde q^{1,2}}$ & $\sigma_{OSD}$ & 
$\sigma_{SSD}$ &$\sigma_{(3\ell+jets)}$ 
& $\sigma_{(3\ell)}$ & $\sigma_{jets}$ \\
\hline 
 (500,500) &{\bf mSUGRA} & 0.2190 & 0.1301 & 0.0316 & 0.00222 & 9.0107 \\
\hline
 (500,500) &{\bf 1000} & 0.2365 & 0.1428 & 0.0351 & 0.00518 & 6.9707 \\
\hline
 (500,500) &{\bf 10000} & 0.2535 & 0.1720 & 0.0608 & 0.02036 & 2.9642  \\
\hline
(500,1000) &{\bf mSUGRA} & 0.1317 & 0.0574 & 0.0160 & 0.00325 & 4.1353 \\
\hline
(500,1000) &{\bf 1000} & 0.0949 & 0.0442 & 0.0067 & \bf{0.00027} & 7.8590  \\
\hline
(500,1000) &{\bf 10000} & 0.2411 & 0.1649 & 0.0577 & 0.02284 & 2.7613  \\
\hline
(1000,1000) &{\bf mSUGRA} & 0.0092 & 0.0069 & 0.0024 & \bf{0.00021} & 0.1921 \\
\hline
(1000,1000) &{\bf 1000} & 0.0052 & 0.0028 & \bf{0.0002} & \bf{0.00024} & 0.5255 \\
\hline
(1000,1000) &{\bf 10000} & 0.0103 & 0.0080 & 0.0035 & \bf{0.00021} & 0.1309  \\
\hline
\end {tabular}
\end {center}
\begin{center}
\begin{tabular}{|c|c|c|c|c|c|c|}
\multicolumn{7}{c}{Table B4 : Cross-sections (pb) for Third family 
scalar non-universality}\\
\multicolumn{7}{c}{$\tan \beta$= 40} \\
\multicolumn{7}{c}{(Figure 4)}\\
\hline
 ($m_{\tilde g}$,$m_{\tilde q^{3}}$) & $m_{\tilde q^{1,2}}$ & $\sigma_{OSD}$ & 
$\sigma_{SSD}$ &$\sigma_{(3\ell+jets)}$ 
& $\sigma_{(3\ell)}$ & $\sigma_{jets}$ \\
\hline 
 (500,500) &{\bf mSUGRA} & 0.2971 & 0.1841 & 0.0515 & 0.00652 & 8.7476 \\
\hline
 (500,500) &{\bf 1000} & 0.2894 & 0.1800 & 0.0563 & 0.01036 & 6.5057 \\
\hline
 (500,500) &{\bf 10000} & 0.2557 & 0.1737 & 0.0617 & 0.01879 & 3.1213  \\
\hline
(500,1000) &{\bf mSUGRA} & 0.1517 & 0.0882 & 0.0206 & 0.00663 & 3.8034 \\
\hline
(500,1000) &{\bf 1000} & 0.0835 & 0.0386 & 0.0068 & \bf{0.00131} & 7.9259  \\
\hline
(500,1000) &{\bf 10000} & 0.2512 & 0.1639 & 0.0509 & 0.02318 & 2.8557  \\
\hline
(1000,1000) &{\bf mSUGRA} & 0.0103 & 0.0076 & 0.0030 & \bf{0.00029} & 0.1947 \\
\hline
(1000,1000) &{\bf 1000} & 0.0069 & 0.0029 & 0.0005 & \bf{0.00026} & 0.5256 \\
\hline
(1000,1000) &{\bf 10000} & 0.0103 & 0.0082 & 0.0038 & \bf{0.00034} & 0.1362  \\
\hline
\end {tabular}
\end {center}

\begin{center}
\begin{tabular}{|c|c|c|c|c|c|c|}
\multicolumn{7}{c}{Table B5 : Cross-sections (pb) for $SO(10)~D$-term non-universality}\\
\multicolumn{7}{c}{$\tan \beta$= 5} \\
\multicolumn{7}{c}{(Figure 6 )}\\
\hline
 $m_{\tilde g}$ & D-term & $\sigma_{OSD}$ & 
$\sigma_{SSD}$ &$\sigma_{(3\ell+jets)}$ 
& $\sigma_{(3\ell)}$ & $\sigma_{jets}$ \\
\hline 
 500 &{\bf mSUGRA} & 0.3720 & 0.2136 & 0.0276 & 0.01380 & 14.1440 \\
\hline
 500 &{\bf 0.5} & 0.3762 & 0.0782 & 0.0349 & 0.01120 & 5.5250 \\
\hline
 500 &{\bf -0.5} & 0.3955 & 0.2402 & 0.0438 & 0.01916 & 12.5007  \\
\hline
 500 &{\bf -1.25} & 0.5638 & 0.3438 & 0.0792 & 0.02999 & 11.6682  \\
\hline
 1000 &{\bf mSUGRA} & 0.0251 & 0.0160 & 0.0046 & \bf{0.00035} & 0.7530 \\
\hline
 1000 &{\bf 0.5} & 0.0221 & 0.0165 & 0.0039 & \bf{0.00040} & 0.7519  \\
\hline
 1000 &{\bf -0.5} & 0.0287 & 0.0173 & 0.0049 & \bf{0.00028} & 0.8043  \\
\hline
 1000 &{\bf -1.25} & 0.0341 & 0.0182 & 0.0056 & \bf{0.00045} & 0.8456  \\
\hline
 1500 &{\bf mSUGRA} & 0.0020 & 0.0012 & 0.0003 & \bf{0.00001} & 0.0702 \\
\hline
 1500 &{\bf 0.5} & 0.0018 & 0.0012 & 0.0003 & \bf{0.00003} & 0.0689 \\
\hline
 1500 &{\bf -0.5} & 0.0024 & 0.0013 & 0.0004 & \bf{0.00012} & 0.0709  \\
\hline
 1500 &{\bf -1.25} & 0.0030 & 0.0014 & 0.0005 & \bf{0.00016} & 0.0720  \\
\hline
\end {tabular}
\end {center}

\begin{center}
\begin{tabular}{|c|c|c|c|c|c|c|}
\multicolumn{7}{c}{Table B6 : Cross-sections (pb) for $SO(10)~ D$-term non-universality}\\
\multicolumn{7}{c}{$\tan \beta$= 40} \\
\multicolumn{7}{c}{(Figure 7 )}\\
\hline
 $m_{\tilde g}$ & D-term & $\sigma_{OSD}$ & 
$\sigma_{SSD}$ &$\sigma_{(3\ell+jets)}$ 
& $\sigma_{(3\ell)}$ & $\sigma_{jets}$ \\
\hline 
 500 &{\bf mSUGRA} & 0.5467 & 0.3360 & 0.0882 & 0.02482 & 13.1779 \\
\hline
 500 &{\bf 0.5} & 0.8111 & 0.4336 & 0.1383 & 0.03341 & 12.7985 \\
\hline
 500 &{\bf -0.5} & 0.5552 & 0.3565 & 0.0898 & 0.02789 & 13.2670  \\
\hline
 500 &{\bf -1.25} & 0.5731 & 0.6209 & 0.1283 & 0.0030 & 23.6538  \\
\hline
1000 &{\bf mSUGRA} & 0.0494 & 0.0303 & 0.0112 & \bf{0.00083} & 0.6668 \\
\hline
 1000 &{\bf 0.5} & 0.0447 & 0.0278 & 0.0097 & \bf{0.00105} & 0.6240  \\
\hline
 1000 &{\bf -0.5} & 0.0505 & 0.0298 & 0.0098 & \bf{0.00073} & 0.6309  \\
\hline
 1000 &{\bf -1.25} & NA & NA & NA & NA & NA  \\
\hline
 1500 &{\bf mSUGRA} & 0.0041 & 0.0023 & 0.0010 & \bf{0.00033} & 0.0532 \\
\hline
 1500 &{\bf 0.5} & 0.0043 & 0.0023 & 0.0009 & \bf{0.00028} & 0.0537 \\
\hline
 1500 &{\bf -0.5} & 0.0026 & 0.0018 & 0.0005 & \bf{0.00005} & 0.0460  \\
\hline
 1500 &{\bf -1.25} & NA & NA & NA & NA & NA \\
\hline
\end {tabular}
\end {center}

\noindent
\begin{center}
\begin{tabular}{|c|c|c|c|c|}
\multicolumn{5}{c}{Table B7 : Cross-sections (pb) for SM background}\\
\hline
 $\sigma_{OSD}$ & $\sigma_{SSD}$ &$\sigma_{(3\ell+jets)}$ 
& $\sigma_{(3\ell)}$ & $\sigma_{jets}$ \\
\hline 
 0.1991 & 0.0900 & 0.0041 & 0.1920 & 2.1015 \\
\hline
\end {tabular}
\end {center}

\end{document}